\documentclass{theoretics}

\usepackage{color}

\title{Testing Distributions of Huge Objects}

\ThCSauthor[affil1]{Oded Goldreich}{oded.goldreich@weizmann.ac.il}[https://orcid.org/0000-0002-4329-135X]
\ThCSauthor[affil2]{Dana Ron}{danaron@tau.ac.il}[https://orcid.org/0000-0001-6576-7200]

\ThCSaffil[affil1]{Faculty of Mathematics and Computer Science, Department of Computer Science, Weizmann Institute of Science, Rehovot, Israel}
\ThCSaffil[affil2]{School of Electrical Engineering, Tel Aviv University, Tel Aviv, Israel}

\ThCSshorttitle{Testing Distributions of Huge Objects}

\ThCSyear{2023}
\ThCSarticlenum{12}
\ThCSreceived{Dec 28, 2022}
\ThCSrevised{May 23, 2023}
\ThCSaccepted{Nov 12, 2023}
\ThCSpublished{Dec 30, 2023}
\ThCSkeywords{Property Testing, Distribution Testing}
\ThCSdoi{10.46298/theoretics.23.12}
\ThCSshortnames{O. Goldreich and D. Ron}
\ThCSthanks{An extended abstract of this work appeared in the proceedings of ITCS 2022.}

\newcommand{\BT}{\begin{theorem}}   \newcommand{\ET}{\end{theorem}}
\newcommand{\BD}{\begin{definition}}   \newcommand{\ED}{\end{definition}}
\newcommand{\BCR}{\begin{corollary}} \newcommand{\ECR}{\end{corollary}}
\newcommand{\BCJ}{\begin{conjecture}} \newcommand{\ECJ}{\end{conjecture}}
\newcommand{\BP}{\begin{proposition}}   \newcommand{\EP}{\end{proposition}}
\newcommand{\BR}{\begin{remark}}   \newcommand{\ER}{\end{remark}}
\newcommand{\BO}{\begin{open}}   \newcommand{\EO}{\end{open}}
\newcommand{\BOB}{\begin{observation}}   \newcommand{\EOB}{\end{observation}}
\newtheorem{Ithm}{Theorem} 
\newcommand{\BIT}{\begin{Ithm}}   \newcommand{\EIT}{\end{Ithm}}
\newcommand{\BL}{\begin{lemma}}   \newcommand{\EL}{\end{lemma}}
\newcommand{\BCM}{\begin{claim}}   \newcommand{\ECM}{\end{claim}}
\newtheorem{techclm}{Claim}[theorem]        %
\newcommand{\Bcm}{\begin{techclm}}   \newcommand{\Ecm}{\end{techclm}}
\newcommand{\BI}{\begin{itemize}}
\newcommand{\EI}{\end{itemize}}
\newcommand{\BDes}{\begin{description}}
\newcommand{\EDes}{\end{description}}

\renewcommand{\eqref}[1]{{\rm Eq.~(\ref{#1})}}
\newcommand{\eqrefs}[2]{{\rm Eq.~(\ref{#1})\&(\ref{#2})}}

\newcommand{\BPF}{\begin{proof}} \newcommand{\EPF}{\end{proof}}
\newcommand{\Bpf}{\begin{subproof}} \newcommand{\Epf}{\end{subproof}}

\newcommand{\myparagraph}[1]{\paragraph{{#1}}}
\newcommand{\mypar}[1]{\subparagraph{{#1}}}

\renewcommand{\sf}{\em}

\newcommand{\prob}{{\rm Pr}}
\newcommand{\Exp}{{\rm E}}

\newcommand{\e}{\epsilon}
\newcommand{\eps}{\e}

\newcommand{\bitset}{\{0,1\}}
\newcommand{\xth}{{\rm th}}

\newcommand{\qc}{{q_{\tt C}}}
\newcommand{\qt}{{q_{\tt T}}}
\newcommand{\qcp}{{q'_{\tt C}}}
\newcommand{\qtp}{{q'_{\tt T}}}

\newcommand{\rst}{{\;\mbox{\rm s.t.}\;}}
\newcommand{\floor}[1]{{\lfloor#1\rfloor}}
\newcommand{\ceil}[1]{{\lceil#1\rceil}}

\newcommand{\eqdef}{\stackrel{\rm def}{=}}

\newcommand{\ov}{\overline}

\newcommand{\tildeO}{{\widetilde{O}}}

\def\niceN{1}
\ifnum\niceN=1
  \font\tenmsb=msbm10 scaled\magstep1
  \font\sevenmsb=msbm7 scaled\magstep1
  \font\fivemsb=msbm5 scaled\magstep1
  \newfam\msbfam
  \textfont\msbfam=\tenmsb
  \scriptfont\msbfam=\sevenmsb
  \scriptscriptfont\msbfam=\fivemsb
  
  \newcommand{\N}{\mathbb N}

\else
  \newcommand{\lndI}{\makebox[0pt][l]{\hspace*{1pt}I}}
  \newcommand{\N}{\mbox{\bf\lndI N}}

\fi
\newcommand{\X}{{\cal X}}
\newcommand{\Y}{{\cal Y}}
\newcommand{\ZZ}{{\cal Z}}
\newcommand{\A}{{\cal A}}

\newcommand{\CP}{{{\tt CP}}}

\newcommand{\poly}{{\rm poly}}

\newcommand{\C}{{\cal C}}

\newcommand{\D}{{\cal D}}

\newcommand{\DP}{{{\cal D}_\Pi}}
\newcommand{\DEC}{{{\tt corr}}}

\newcommand{\STOC}{ACM Symposium on the Theory of Computing}
\newcommand{\FOCS}{IEEE Symposium on Foundations of Computer Science}
\newcommand{\ITCS}{Innovations in Theoretical Computer Science}

\newcommand{\SICOMP}{SIAM Journal on Computing}

\newcommand{\cP}{P}
\newcommand{\cQ}{Q}
\newcommand{\Dom}{\Omega}

\newcommand{\InEq}{{\rm InEq}}
\newcommand{\DHO}{\textsc{DoHO}}
\newcommand{\DHOm}{{\DHO\ model}}
\newcommand{\ccode}{{\tt cc}}
\newcommand{\Dper}{{\mathcal{D}^{\rm per}_{\eta,\delta}}}
\newcommand{\DperPi}{{\mathcal{D}^{{\rm per},\Pi}_{\eta,\delta}}}
\newcommand{\Dcyc}{{\mathcal{D}^{\rm cyc}}}
\newcommand{\Dcycu}{{\mathcal{D}^{\rm Ucyc}}}
\newcommand{\DcycR}{{\mathcal{D}^{{\rm cyc}_I}}}
\newcommand{\Diso}{{\mathcal{D}^{\rm iso}}}

\begin{document}
\maketitle

\begin{abstract}
We initiate a study of a new model of property testing
that is a hybrid of testing properties of distributions
and testing properties of strings. Specifically,
the new model refers to testing properties of distributions,
but these are distributions over huge objects (i.e., very long strings).
Accordingly, the model accounts for the total number of local probes
into these objects (resp., queries to the strings)
as well as for the distance between objects (resp., strings),
and the distance between distributions is defined
as the earth mover's distance with respect to
the relative Hamming distance between strings.

We study the query complexity of testing in this new model,
focusing on three directions. First, we try to relate the
query complexity of testing properties in the new model
to the sample complexity of testing these properties
in the standard distribution-testing model.
Second, we consider the complexity of testing properties
that arise naturally in the new model (e.g., distributions
that capture random variations of fixed strings).
Third, we consider the complexity of testing properties that
were extensively studied in the standard distribution-testing model:
Two such cases are uniform distributions
and pairs of identical distributions,
where we obtain the following results.

\begin{itemize}
\item
Testing whether a distribution over $n$-bit long strings
is uniform on some set of size $m$ can be done
with query complexity ${\widetilde O}(m/\epsilon^3)$,
where $\epsilon>(\log_2m)/n$ is the proximity parameter.
\item
Testing whether two distribution over $n$-bit long strings
that have support size at most $m$ are identical can be done
with query complexity ${\widetilde O}(m^{2/3}/\epsilon^3)$.
\end{itemize}
Both upper bounds are quite tight; that is,  for $\epsilon=\Omega(1)$,
the first task requires $\Omega(m^c)$ queries
for any $c<1$ and $n=\omega(\log m)$,
whereas the second task requires $\Omega(m^{2/3})$ queries.
Note that the query complexity of the first task is higher
than the sample complexity of the corresponding task
in the standard distribution-testing model, whereas in
the case of the second task the bounds almost match.
\end{abstract}


\section{Introduction}
In the last couple of decades,
the area of property testing has attracted much attention
(see, e.g., a recent textbook~\cite{G:pt}).
Loosely speaking, property testing typically refers to sub-linear time
probabilistic algorithms for {\em approximate decision}\/ making;
that is, distinguishing between objects that have a predetermined property
and ones that are far from any object having this property.
Such algorithms, called testers, obtain local views of the object
by making adequate queries; that is, the object is modeled as a function
and  testers get oracle access to this function
(and thus may be expected to work in time that
is sub-linear in the size of the object).

The foregoing description fits much of the research
in the area (see~\cite[Chap.~1-10]{G:pt}),
but not the part that deals with testing properties of distributions
(aka {\em distribution testing},
see~\cite[Chap.~11]{G:pt} and~\cite{C:survey,C:survey2}).
In this context, a tester get samples from the tested distribution
and sub-linearity means sub-linearity
in the size of the distribution's domain.
\footnote{This is the most standard and well studied model
of testing properties of distributions.
For a discussion of other models (e.g., providing the algorithm
with the weight of any domain element of its choice)
see~\cite[Part IV]{C:survey}.}
Each element 
in the domain is considered to be small, and is
assumed to be processed at unit time,
and the distance between distributions
is defined as the total variation distance.

In this work
we consider distributions over sets of huge (or very large) objects,
and aim at complexities that are sublinear in the size of these objects.
As an illustrative example, think of the distribution of DNA-sequences
in a large population. We wish to sample this distribution and
{\em query each sampled sequence at locations of our choice
rather than read the entire sample}.

One key issue is the definition of the distance between
such distributions (i.e., distributions of huge objects).
A natural choice, which we use, is 
the
{\em earth mover's distance}\/ under the (relative) Hamming distance.
Under this combined measure, the distance between distributions reflects
the probability mass that ``needs to be moved'' when weighted according to
the Hamming distance between strings (see Definition~\ref{dist:def}).
Thus, a probability mass of $p$ moved from a string~$x$
(in one distribution) to a string~$y$ (in the other distribution)
contributes $p\cdot\delta$ units to the distance between the distributions,
where $\delta$ is the relative Hamming distance between $x$ and $y$.

Recall that the basic paradigm of distribution testing postulates that,
in many settings, it suffices to distinguish distributions that satisfy
the property (of interest) from ones that are far from satisfying it;
that is, it postulates that a notion of approximate decision is meaningful.
In fact, this basic paradigm underlies all of statistics.
Here, we suggest that, in the case of distributions on huge objects,
a distance measure that is more refined than total variation distance
is more adequate. Specifically,
{\em the suggested distance measure accounts for the distance between
elements of the distributions' domain rather than viewing them as atoms}.
Indeed, this suggestion goes hand in hand with a more refined notion
of inspecting an element of the domain (or of the sample);
it allows to partially inspect an element rather than forcing us
to either look at the entire element or not look at it at all.

We mention that, while we follow the archetypical convention
of property testing in using the Hamming distance as the distance
between strings (i.e., objects), our framework allows to use
any other distance between strings as a basis for defining distance
between distributions on huge objects (i.e., long strings).
For example, one may consider the edit distance between strings.
Indeed, standard distribution testing is obtained as a special case
by considering the equality function as a distance between strings.

To illustrate the flavor of our study, consider the problem
of testing whether a given distribution over $n$-bit strings
has support size $m$ (resp., is uniform over $m$ strings).
When $n$-bit strings are considered as atoms,
the results of distribution testing apply and assert that,
when ignoring the (polynomial) dependence on the proximity parameter,
the sample complexity of this problem
is $\Theta(m/\log m)$~\cite{VV,VV-jacm}
(resp., $\Theta(m^{2/3})$~\cite{BC}).
But the resulting tester requires entirely reading each of the samples
(i.e., an $n$-bit long string),
which means reading $\Theta(nm/\log m)$ bits.
The question we address is whether the query complexity
may be reduced
(under the definitions outlined above and detailed next).
(Jumping ahead, the answer is positive:
Both properties can be tested using $\tildeO(m)$ queries.)

\subsection{The new model}
We consider properties of distributions over sets of objects
that are represented by $n$-bit long strings
(or possibly $n$-symbol long sequences);
that is, each object has size $n$.
(In Section~\ref{tuples:sec} this will be extended
to properties of tuples of distributions.)
Each of these objects is considered huge,
and so we do not read it in full
but rather probe (or query) it at locations of our choice.
Hence, the tester is an algorithm that may ask for few samples,
and queries each sample at locations of its choice.
This is modeled as getting oracle access to several oracles,
where each of these oracles is selected independently 
according to the tested distribution
(see Definition~\ref{dist:test-one.def}).
We shall be mainly interested in the total number of queries
(made into these samples), whereas the number of samples
will be a secondary consideration.

The distance between such distributions, 
$\cP$ and $\cQ$ (over the same domain $\Dom=\bitset^n$),
is defined as the {\em earth mover's distance under the Hamming measure};
that is, the cost of transforming the distribution $\cP$
to the distribution $\cQ$, where the cost of transforming
a string $x$ to a string~$y$ equals their relative Hamming distance.
The minimum cost of such a transformation (of $P$ to $Q$) is captured
by \eqref{EMD-via-LP:eq}; an alternative formulation appears
in \eqref{general:cost-of-move:eq} (in Appendix~\ref{emd-tvd:apdx}).

\BD{\rm(Distance between distributions over huge objects):}
\label{dist:def}
For two strings $x,y\in \bitset^n$,
let $\Delta_H(x,y)$ denote the relative Hamming distance between them;
that is,
\begin{equation}\label{Hamming-distance:eq}
\Delta_H(x,y) = \frac{1}{n}\cdot |\{i\in [n]\;:\; x_i \neq y_i\}|.
\end{equation}
For two distributions $\cP,\cQ : \Dom \to [0,1]$, where $\Dom=\bitset^n$, the
{\sf earth mover's distance under the Hamming measure} between $\cP$ and $\cQ$, is the optimal value of
the following linear program {\rm(in the variables $w_{x,y}$'s)}:
\begin{equation} 
\min_{\begin{array}{ll}
   \scriptstyle \forall x,y\in \Dom:&
   \scriptstyle w_{x,y}\geq 0 \\
   \scriptstyle\forall x\in \Dom: &
   \scriptstyle\sum_{y\in \Dom} w_{x,y} = \cP(x) \\
   \scriptstyle \forall y\in \Dom: &
   \scriptstyle \sum_{x\in \Dom} w_{x,y} = \cQ(y)
    \end{array}}
       \left\{\sum_{x,y\in \Dom} w_{x,y}\cdot \Delta_H(x,y)\right\}_{.}
\label{EMD-via-LP:eq}
\end{equation}
We say that {\sf $\cP$ is $\e$-close to $\cQ$}
if the optimal value of the linear program is at most $\e$;
otherwise, we say that {\sf $\cP$ is $\e$-far from~$\cQ$}.
\ED
As stated above, Definition~\ref{dist:def} represents
the earth mover's distance with respect to the relative Hamming distance between (binary) strings.
Indeed, the earth mover's distance between distributions over a domain $\Dom$
is always defined on top of a distance measure that is associated
with $\Dom$.
It is well known that the earth mover's distance with respect to
the inequality function (i.e., $\InEq(x,y)=1$ if $x\neq y$ and $\InEq(x,x)=0$)
coincides with 
the variation distance (between these distributions).
That is, if we replace 
the distance $\Delta_H(x,y)$ with $\InEq(x,y)$ in Definition~\ref{dist:def},
then we get 
the variation distance between $\cP$ and $\cQ$
(see Appendix~\ref{emd-tvd:apdx}).
Furthermore, $\Delta_H(x,y)\leq\InEq(x,y)$ always holds.
Hence, the distance between pairs of distributions
according to Definition~\ref{dist:def}
is upper-bounded by the total variation distance between them.

\medskip
\noindent
Hence, throughout this work,
we shall be considering three distance measures:

\smallskip
\begin{enumerate} 
\item The {\em distance between distributions}\/
as defined above (i.e., in Definition~\ref{dist:def}).
When we say that distributions are ``close'' or ``far''
we refer to this notion.
\item The {\em total variation distance between distributions}.
In this case, we shall say that the distributions are ``TV-close''
or ``TV-far'' (or $\e$-TV-close/far).
\item The {\em relative Hamming distance between strings},
which we denoted by $\Delta_H(\cdot,\cdot)$.
In this case, we shall say that the strings are ``H-close''
or ``H-far'' (or $\e$-H-close/far).
\end{enumerate} 
Referring to Definition~\ref{dist:def} and to machines
that have access to multiple oracles, we present the following definition
of testing distributions on huge objects.



\BD{\rm(Testing properties of distributions on huge objects (the \DHOm)):}
\label{dist:test-one.def}
Let ${\cal D}=\{{\cal D}_n\}_{n\in\N}$ be a property of distributions
such that ${\cal D}_n$ is a set of distributions over $\bitset^n$,
and let $s:\N\times(0,1]\to\N$.
A {\sf tester}, denoted $T$, {\sf of sample complexity $s$}
{\sf for the property $\cal D$} is a probabilistic machine that,
on input parameters $n$ and $\e$,
and oracle access to a sequence of $s=s(n,\e)$ samples
drawn from an unknown distribution 
$\cP : \bitset^n \to [0,1]$
outputs a verdict {\rm(``accept'' or ``reject'')}
that satisfies the following two conditions.

\smallskip
\begin{enumerate} 
\item{\em The tester accepts distributions that belong to $\cal D$:}
If $\cP$ is in ${\cal D}_n$, then
$$\prob_{x^{(1)},\dots,x^{(s)}\sim \cP}[T^{x^{(1)},\dots,x^{(s)}}(n,\e)\!=\!1]
   \geq2/3,$$
where $x^{(1)},\dots,x^{(s)}$ are drawn
independently from the distribution $\cP$,
and the probability is taken ver these draws
as well as over the internal random choices of $T$.
\item{\em The tester rejects distributions that are far from $\cal D$:}
If $\cP$ is $\e$-far from ${\cal D}_n$
{\rm(i.e., $\cP$ is $\e$-far from any distribution in ${\cal D}_n$
(according to Definition~\ref{dist:def}))}, then
$$\prob_{x^{(1)},\dots,x^{(s)}\sim \cP}[T^{x^{(1)},\dots,x^{(s)}}(n,\e)\!=\!0]
  \geq2/3,$$
where $x^{(1)},\dots,x^{(s)}$ and the probability space
are as in the previous item.
\end{enumerate} 
We say that $q:\N\times(0,1]\to\N$ is
the {\sf query complexity of $T$} if $q(n,\e)$ is the maximum number
of queries that $T$ makes on input parameters $n$ and $\e$.
If the tester accepts every distribution in $\cal D$ with probability~1,
then we say that it has {\sf one-sided error}.
%
\ED
We may assume, without loss of generality,
that the tester queries each of its samples,
and that it never makes the same query twice.
Hence, $q(n,\e)\in[s(n,\e),s(n,\e)\cdot n]$.

The {\sf sample} (resp., {\sf query})
{\sf complexity of testing} the property $\cal D$ (in the \DHOm)
is the minimal sample (resp., query) complexity of a tester
for $\cal D$ (in the \DHOm).
Note that the tester achieving the minimal sample complexity
is not necessarily the one achieving the minimal query complexity.
As stated before, we shall focus on minimizing the query complexity,
while using the sample complexity as a yardstick.

\paragraph{Generalization.}
The entire definitional treatment can be extended to $n$-long
sequences over an alphabet $\Sigma$, where above
(in Definitions~\ref{dist:def} and~\ref{dist:test-one.def})
we used $\Sigma=\bitset$.

\subsection{The standard notions of testing as special cases
(and other observations)}
We first observe that both the standard model of property testing
(of strings) and the standard model of distribution testing
are special cases of Definition~\ref{dist:test-one.def}.
\BDes
\item[{\bf Standard property testing (of strings)}:]
Specifically, we refer to testing properties of $n$-bit strings
(equiv., Boolean functions over $[n]$).

This special case corresponds to trivial distributions,
where each distribution is concentrated on a single $n$-bit long string.
Hence, a standard tester of query complexity $q$ can be
viewed as a tester in the sense of Definition~\ref{dist:test-one.def}
that has sample complexity~1 and query complexity~$q$.

\item[{\bf Standard distribution testing}:]
Specifically, we refer to testing distributions over $\Sigma$.

This special case corresponds to the case of $n=1$,
where each distribution is over $\Sigma$.
Hence, a standard distribution tester of sample complexity $s$ can be
viewed as a tester in the sense of Definition~\ref{dist:test-one.def}
that has sample complexity~$s$ and query complexity $q=s$.
Indeed, here we used the generalization of the definitional
treatment to sequences over $\Sigma$. The basic version,
which refers to bit sequences, can be used too (with a small overhead).%
\footnote{Specifically, we consider a good error correcting
code $C:\Sigma\to\bitset^n$ such that $n=O(\log|\Sigma|)$;
that is, $C$ has distance $\Omega(n)$. In this case,
the total variation distance between distributions over codewords
is proportional to their distance according to Definition~\ref{dist:def},
whereas the query complexity is at most $n=O(\log|\Sigma|)$ times
the sample complexity.
The same effect can be obtained by using larger $n$'s,
provided we use locally testable and correctable codes.}
\EDes
Needless to say,
the point of this paper is going beyond these standard notions.
In particular, we seek testers (for the \DHOm)
with query complexity $q(n,\e)=o(n)\cdot s(n,\e)$,
where $s(n,\e)>1$ is the sample complexity in the \DHOm.
Furthermore, our focus is on cases in which $s(n,\e)$
is relatively small (e.g., $s(n,\e)=\poly(n/\e)$
and even $s(n,\e)=o(n)\cdot\poly(1/\e)$),
since in these cases a factor of $n$ matters more.

We mention that the sample complexity in the \DHOm\
is upper-bounded by the sample complexity
in the standard distribution-testing model.
This is the case because the distance between pairs of distributions
according to Definition~\ref{dist:def}
is upper-bounded by the total variation distance between them
(see the discussion following Definition~\ref{dist:def}).
We state this observation for future reference.

\begin{observation}{{\em (On the sample complexity of testing distributions in two models):}}
\label{obs:samp-samp}
The sample complexity of testing a property $\D$ of distributions
over $\bitset^n$ in the \DHOm\ is upper-bounded by
the sample complexity of testing $\D$
in the standard distribution-testing model.
\end{observation}
We mention that, for some properties $\D$,
the sample complexity in the \DHOm\
may be much lower than in the standard distribution-testing model,
because in these cases the distance measure in the \DHOm\
is much smaller than the total variation distance.%
\footnote{An obvious case is one in which testing distributions
is trivial (in the \DHOm). This is the case for the set
of all distributions that are supported by a set of strings $\Pi$
such that any string is H-close to $\Pi$.
Specifically, if every $n$-bit long string is $\e$-H-close
to $\Pi\subseteq\bitset^n$ and $\D$ is the set of distributions
that contain every distribution that is supported by $\Pi$,
then every distribution is $\e$-close to $\D$.
On the other hand, testing $\D$ in the standard model may be non-trivial.
Additional examples are presented in Section~\ref{triviality:sec}.}
%
Needless to say, this is not true in general,
and we shall focus on cases in which the two
sample complexities are closely related.
In other words, we are not interested in the possible gap
between the sample complexities (in the two models),
although this is a legitimate question too,
but rather in the query complexity in the \DHOm.
Furthermore, we are willing to increase the sample complexity
of a tester towards reducing its query complexity in the \DHOm\
(e.g., see our tester for uniformity).

\subsection{Our Results}\label{subsec:res}
We present three types of results.
The first type consists of  general results that relate
the query complexity of testing in the \DHO\ model
to the query and/or sample complexity of related properties
in the standard (distribution and/or string) testing models.
The second type consists of results for properties that have been
studied (some extensively) in the standard distribution-testing model.
The third type consists of  results for
new properties that arise naturally in the \DHO\ model.

\subsubsection{Some general bounds on the query complexity of testing
in the {\DHO} model}
A natural class of properties of distribution over huge objects
is the class of all distributions that are supported by strings
that have a specific property (of strings).
That is, for a property of bit strings $\Pi=\{\Pi_n\}_{n\in\N}$
such that $\Pi_n\subseteq\bitset^n$,
let $\D_\Pi=\{{\D}_n\}_{n\in\N}$ such that $\D_n$ denotes the set
of all distributions that have a support that is subset of $\Pi_n$.
We observe that the query complexity of testing
the set of distributions $\D_\Pi$ (in the \DHO\ model) is related to
the query complexity of testing the set of strings $\Pi$
(in the standard model of testing properties of strings).

\BT{\rm(From testing strings for membership in $\Pi$
to testing distributions for membership in $\D_\Pi$):}
\label{support:ithm}
If the query complexity of testing $\Pi$ is $q$,
then the query complexity of testing $\D_\Pi$ in the \DHOm\
is at most~$q'$ such that $q'(n,\e)=\tildeO(1/\e)\cdot q(n,\e/2)$.
\ET
While the proof of Theorem~\ref{support:ithm} is simple,
we believe it is instructive towards getting familiar with the \DHOm.
We thus include it here, while mentioning that some ramifications
of it appear in Appendix~\ref{general-equality:apdx}.
\medskip

\BPF
The main observation is that if the tested distribution $\cP$
(whose domain is $\bitset^n$)
is $\e$-far from $\D_n$ (according to Definition~\ref{dist:def}),
then, with probability at least $\e/2$,
an object $x$ selected according to $\cP$ is $\e/2$-H-far from $\Pi_n$.
Hence, with high constant probability,
a sample of size $O(1/\eps)$ will contain at least one string
that is $\e/2$-H-far from $\Pi_n$.
If we have a one-sided error tester $T$ for $\Pi$,
then we can detect this event (and reject) by running $T$
(with proximity parameter $\e/2$) on each sampled string.
If we only have a two-sided error tester for $\Pi$,
then {\em we invoke it $O(\log(1/\e))$ times on each sample},
and reject if the majority rule regarding any of these samples is rejecting.
Hence, in total we make $O(\e^{-1}\log(1/\e))\cdot q(n,\e/2)$ queries.
\EPF

\paragraph{An opposite extreme.}
Theorem~\ref{support:ithm} applies to any property $\Pi$ of strings
and concerns the set of \emph{all} distributions that are
supported by $\Pi$ (i.e., all distributions $P$ that
satisfy $\{x\!:\!P(x)\!>\!0\}\subseteq\Pi$).
Hence, Theorem~\ref{support:ithm} focuses on the support
of the distributions and pays no attention to all other~aspects
of the distributions. The other extreme is to focus on properties
of distributions that~are invariant under relabeling of the strings
(i.e., label-invariant properties of distributions).%
\footnote{Recall that a property of distributions over $\bitset^n$
is called {\sf label-invariant} if,
for every bijection $\pi:\bitset^n\to\bitset^n$
and every distribution $\cP$,
it holds that $\cP$ is in the property
if and only if $\pi(\cP)$ is in the property,
where $\cQ=\pi(\cP)$ is the distribution defined
by $\cQ(y) = \cP(\pi^{-1}(y))$.
We mention that label-invariant properties of distributions
are often called symmetric properties.}
We consider several such specific properties in Section~\ref{subsubsec:prev},
but in the current section we seek more general results.
Our guiding question is the following.

\BO{\em(A key challenge, relaxed formulation):}%
\footnote{Our formulation uses $s(n,\e/2)$ rather than $s(n,\e)$
in order to allow some slackness in a possible transformation.
Less relaxed formulations may require
query complexity $\tildeO(s(n,\e/2)/\e)$ or even $O(s(n,\e))$.
On the other hand, one may ease the requirement
by comparing the query complexity in the \DHOm\
to the sample complexity in the standard model.}
\label{label-invariant:open}
For which label-invariant properties of distributions
does it hold that testing them in the \DHOm\
has query complexity $\poly(1/\e)\cdot\tildeO(s(n,\e/2))$,
where $s$ is the sample complexity of testing them in the \DHOm?
\EO
Jumping ahead, we mention that
in Section~\ref{subsubsec:prev}
we identify two label-invariant properties
for which the relation between the query complexity
and the sample complexity is as sought in 
Problem~\ref{label-invariant:open},
and one for which this relation does not hold.
More generally, we show that a relaxed form of 
such a relation
(in which $s$ is the sample complexity in the standard model)
is satisfied for any property that is closed under mapping,
where a property of distribution~$\D$ is {\sf closed under mapping}
if, for every distribution $\cP:\bitset^n\to[0,1]$ in $\D$
and every $f:\bitset^n\to\bitset^n$,
it holds that $f(\cP)$ is in $\D$,
where $\cQ=f(\cP)$ is the distribution defined by $\cQ(y) = \cP(f^{-1}(y))$.

\BT{\em(Testing distributions that are closed under mapping
(see Theorem~\ref{closed-mapping:thm})):}
\label{closed-mapping:ithm}
Suppose that $\D=\{\D_n\}$ is testable
with sample complexity $s(n,\e)$ in the standard model,
and that each~$\D_n$ is closed under mapping.
Then, $\D$ is testable in the \DHOm\
with query complexity $\tildeO(\e^{-1}\cdot s(n,\e/2))$.
\ET
Recall that a tester of sample complexity $s$
in the standard distribution-testing model constitutes
a tester of sample complexity $s$ in the \DHOm,
alas this tester has query complexity $n\cdot s$
(whereas our focus is on the case
that $n\gg\poly(\e^{-1}\log s(n,\e/2))$).
We wonder whether a result similar to Theorem~\ref{closed-mapping:ithm}
holds when $s$ is the sample complexity in the \DHOm.%
\footnote{Such a result was wrongly claimed in Revision~1
of our ECCC TR21-133 (and in the extended abstract
that appeared in the proceedings of {\em ITCS22}).
Partial progress towards such a result
is presented in Appendix~\ref{closed-mapping:apdx}.}


A middle ground between properties that contain all distributions
that are supported by a specific set of strings
and label-invariant properties of distributions
is provided by properties of distributions that
are label-invariant only on their support,
where the {\sf support of a property of distributions}
is the union of the supports of all distributions in this property.
That is, for a property $\D_n$ of distributions over $n$-bit strings,
we say that $\D_n$ is {\sf label-invariant over its support}
if, for every bijection $\pi:\bitset^n\to\bitset^n$
that {\em preserves the support of $\D_n$}\/
(i.e., $x$ is in the support if and only if $\pi(x)$ is in the support),
it holds that the distribution $\cP:\bitset^n\to[0,1]$ is in $\D_n$
if and only if $\pi(\cP)$ is in $\D_n$.
Indeed, generalizing Problem~\ref{label-invariant:open},
one may ask 

\BO{\em(A more general challenge):}
\label{label-invariant-on-support:open}
For which properties of distributions that are
label-invariant over their support
does it hold that testing them in the \DHOm\
has query complexity $\poly(1/\e)\cdot\tildeO(s(n,\e/2)\cdot(q(n,\e/2)+1))$,
where $s$ is the sample complexity of testing them in the \DHOm\
and $q$ is the query complexity of testing their support?
\EO
The next theorem identifies a sufficient condition for a positive answer.
Specifically, it requires that the support of the property, denoted $S$,
has a (relaxed) {\em self-correction}\/ procedure of query complexity~$q$.
We mention that such procedures may exist only in case the
strings in $S$ are pairwise far apart.
Loosely speaking,
on input $i\in[n]$ and oracle access to an $n$-bit string $x$,
the self-correction procedure is required to return $x_i$ if $x\in S$,
to reject if $x$ is far from $S$,
and otherwise it should either reject or return the $i^\xth$ bit of
the string in $S$ that is closest to $x$.

\BT{\em(Self-correction-based testers in the \DHOm,
loosely stated (see Theorem~\ref{li+sc:thm})):}
\label{tentative:ithm}
Let $\D$ be a property of distributions over bit strings
that is label-invariant over its support.
Then, ignoring polylogarithmic factors,
the query complexity of testing $\D$ in the \DHO\ model
is upper-bounded by the product of the sample complexity
of testing $\D$ in the standard model and the query
complexity of testing and self-correcting the support of $\D$.
\ET
One natural example to which Theorem~\ref{tentative:ithm}
is applicable is a set of all distributions that each has
a support that contains ``few'' low-degree multi-variate polynomials;
specifically, for support-size bound $s(n)$ and the degree bound $d(n)$,
we get query complexity $\poly(d(n)/\e)\cdot\tildeO(s(n))$.
Note that the dependency on the support size is quite tight
in light of Theorem~\ref{lb:support+uniform:ithm}
(see the proof of Proposition~\ref{lb:m-support:obs}).

\subsubsection{Testing previously studied properties of distributions}
\label{subsubsec:prev}
Turning back to label-invariant properties of distributions,
we consider several such properties that were studied previously
in the context of the standard distribution-testing model.
Specifically, we consider the properties of
having bounded support size (see, e.g.,~\cite{RRSS}),
being uniform over a subset of specified size (see, e.g.,~\cite{BC}),
and being $m$-grained (see, e.g.,~\cite{GR:gr}).%
\footnote{A distribution $P:\bitset^n\to[0,1]$ is called {\sf $m$-grained}
if any $n$-bit string appears in it with probability that is
a multiple of $1/m$; that is, for every $x\in\bitset^n$
there exists an integer $m_x$ such that $P(x)=m_x/m$.}

\BT{\em(Testers for support size, uniformity, and $m$-grained in the \DHOm\
(see Corollary~\ref{support+uniform:cor})):}
\label{support+uniform:ithm}
For any $m$, the following properties of distributions over $\bitset^n$
can be tested in the \DHO\ model using $\poly(1/\e)\cdot\tildeO(m)$ queries:

\smallskip
\begin{enumerate} 
\item All distributions having support size at most $m$.
\item All distributions that are uniform over some set of size $m$.
\item All distributions that are $m$-grained.
\end{enumerate} 
\ET
Theorem~\ref{support+uniform:ithm} is proved
by using Theorem~\ref{closed-mapping:ithm}.
The foregoing upper bounds are quite tight
(see Theorem~\ref{lb:support+uniform:ithm}).
They also provide positive and negative cases
regarding Problem~\ref{label-invariant:open}
(see discussion following Theorem~\ref{lb:support+uniform:ithm}).

\BT{\em(Lower bounds on testing support size, uniformity,
and $m$-grained in the \DHOm\
(see Propositions~\ref{lb:m-support:obs},~\ref{new-lb:m-uniformity:prop}
and~\ref{new-lb:m-grained:prop})):}
\label{lb:support+uniform:ithm}

\smallskip
\begin{enumerate} 
\item
For every $m\leq2^{n-\Omega(n)}$,
testing whether a distribution over $\bitset^n$
has support size at most $m$ requires $\Omega(m/\log m)$ samples.
\item
For every constant $c<1$ and $m\leq{n}$,
testing whether a distribution over $\bitset^n$
is uniform over some subset of size $m$ requires $\Omega(m^c)$ queries.
\item
For every constant $c<1$ and $m\leq2^{n-\Omega(n)}$,
testing whether a distribution over $\bitset^n$
is $m$-grained requires $\Omega(m^c)$ samples.
\end{enumerate} 
\ET
Note that Parts~1 and~3 assert
lower bounds on the {\em sample complexity}\/ in the \DHOm, which imply
the same lower bounds on the {\em query complexity}\/ in this model.
Combining the first part of Theorems~\ref{support+uniform:ithm}
and~\ref{lb:support+uniform:ithm} yields a property that satisfies
the requirement of Problem~\ref{label-invariant:open};
that is, the query complexity in the \DHOm\ is closely
related to the sample complexity (in this model).
On the other hand, combining Part~2 of Theorem~\ref{lb:support+uniform:ithm}
with the tester of~\cite{BC,DKS} yields a property that does not satisfy the
requirement in Problem~\ref{label-invariant:open},
since this tester uses $O(m^{2/3}/\e^2)$ samples
(even in the standard distribution-testing model).%
\footnote{We mention that in~\cite{BC,DKS} the complexity bound is
stated in terms of the second and third norms of the tested distribution,
which can be roughly approximated by the number of samples required
for seeing the first 2-way and 3-way collisions.
To obtain complexity bounds in terms of $m$,
we can take $O(m^{2/3})$ samples and reject if
no 3-way collision is seen (ditto for not seeing a 2-way
collision among the first $O(m^{1/2})$ samples).}

\paragraph{Tuples of distributions.}
In Section~\ref{tuples:sec} we extend the \DHO\ model to testing
tuples (e.g., pairs) of distributions, and consider
the archetypical problem of testing equality of distributions
(cf.~\cite{BFRSW,BFRSW-long}).
In this case, we obtain another natural property that satisfies
the requirement of Problem~\ref{label-invariant:open}.

\BT{\em(A tester for equality of distributions
(see Theorem~\ref{equal-distributions:thm})):}
\label{equal-distributions:ithm}
For any $m,n\in\N$ and $\e>0$,
given a pair of distributions over $\bitset^n$
that have support size at most $m$,
we can distinguish between
the case that the distributions are identical
and the case that they are $\e$-far from one another
{\em(according to Definition~\ref{dist:def})}
using $\tildeO(m^{2/3}/\e^3)$ queries and $O(m^{2/3}/\e^2)$ samples.
\ET
We note that $m^{2/3}/\e^2$ is
a proxy for $\max(m^{2/3}/\e^{4/3},m^{1/2}/\e^2)$,
which is a lower bound on the sample complexity of testing
this property in the standard distribution-testing model~\cite{V}.
This lower bound can be extended to the \DHOm.
Hence, in this case, the query complexity in the \DHOm\
is quite close to the sample complexity in this model.

\subsubsection{Distributions as variations of an ideal object}
\label{subsubsec:vars}
A natural type of distributions over huge objects arises
by considering random variations of some ideal objects.
Here we assume that we have no access to the ideal object,
but do have access to a sample of random variations of this object,
and we may be interested both in properties of the ideal object
and in properties of the distribution of variations.
We consider three types of such variations,
and provide testers for the corresponding properties.

\paragraph{Noisy versions of a string, where we bound the noise level.}
Specifically, we consider a noise model in which each bit
may be flipped with some bounded probability $\eta<1/2$
such that the resulting string is at relative Hamming distance
at most $\delta$ from the original string.
In this case it is easy to recover bits of the original string,
and test that the noisy versions respect the predetermined noise model.
Furthermore, we can test properties of the original string
with rather small overhead.

\paragraph{Random cyclic-shifts of a string.}
Specifically, we consider distributions obtained by applying
arbitrary random cyclic-shift to a fixed $n$-bit string
as well as distributions obtained by shifting a fixed $n$-bit string
by a uniformly distributed number of positions.
The first property (i.e., random cyclic-shifts)
is tested using $O(1/\e)$ samples and $\tildeO({\sqrt n}/\e)$ queries
(see Theorem~\ref{random-shifts:thm}),
whereas for testing the second property
(i.e., uniformly random cyclic-shifts)
we use $\tildeO(n^{2/3}/\e)$ queries (see Theorem~\ref{uniform-shifts:thm}).
The first tester uses a tester of cyclic-shifts
(i.e., given two strings, the tester checks whether
one is a cyclic shift of the other),
whereas the second testing problem is reduced
to testing equality between distributions
(as addressed in Theorem~\ref{equal-distributions:ithm}).
In the latter case, the two distribution that
we test for equality are the given distribution
and the distribution defined by a uniform random shift
of a single sample obtained from the given distribution.
(The reduction is requires a generalization
of Theorem~\ref{equal-distributions:ithm} to the case that
only the support size of one of the distributions is bounded.)

\paragraph{Random isomorphic copies of a graph
(represented by its adjacency matrix).}
Specifically, here the $n$-bit string represents
the adjacency matrix of an $\sqrt n$-vertex graph,
and the tested property is an arbitrary distribution
over isomorphic copies of some fixed graph.
In this case we get a tester of query complexity
$\poly(1/\e)\cdot\tildeO(n^{5/8})$,
by employing a known tester for graph isomorphism~\cite{FM}.

\subsection{Orientation and Techniques}
\label{conventions:sec}
As stated upfront, we seek testers that sample the distribution
but do not read any of the samples entirely
(but rather probe some of their bits).
In other words, we seek testers for distributions over $n$-bit strings
of query complexity that is $o(n)$ times the sample complexity
of testing these distributions in the standard model.

In general, our proofs build on first principles,
and are not technically complicated.
Rather, each proof is based on one or few observations,
which, once made, lead the way to obtaining the corresponding result.
Hence, the essence of these proofs is finding
the right point of view from which the observations arise.

\paragraph{Conventions.}
It is often convenient to treat distributions as random variables;
that is, rather than referring to the distribution $\cP:\Omega\to[0,1]$
we refer to the random variable $X$ such that $\prob[X\!=\!x]=\cP(x)$.
Recall that $\e$ denotes the proximity parameter
(for the testing task).
Typically, the upper bounds specify the dependence on $\e$,
whereas the lower bound refer to some fixed $\e=\Omega(1)$.

\subsubsection{Our testers}
Some of our testers refer to label-invariant properties
(see Theorems~\ref{closed-mapping:ithm} and~\ref{support+uniform:ithm}).
The key observation here is that, in this case,
it suffices to determine which samples are equal
and which are different. 
Furthermore, considering close samples as if they were equal
does not really create a problem,
because we are working under Definition~\ref{dist:def}.
Hence, checking approximate equality between strings suffices,
and it can be performed
by probing few random locations in the strings.

Unfortunately, the analysis does not reduce to the foregoing comments,
because we cannot afford to consider all strings in the
({\em a priori}\/ unknown) support of the tested distribution;
that is, we cannot argue that the collision pattern among
the projections (at few random locations) correctly represent the
distances between all pairs of strings in the support of the distribution.
Instead, the analysis refers to the empirical distribution
defined by a sequence of samples; that is, we show that
the collision pattern among these projections correctly represent
the distances between all pairs of {\em sampled}\/ strings.
This is done by considering a mental experiment in which
the tested distribution is replaced by an imaginary distribution
that is close to it. This strategy is illustrated next.

\paragraph{Illustration: The case of testing support size.}
Suppose that we wish to test whether a distribution $X$
has support size at most $m$.
Our starting point is the observation that
{\em if $X$ is $\e$-far from having support size at most $m$}\/
(according to Definition~\ref{dist:def}),
{\em then taking $s=O(m/\e)$ samples from $X$ yields}\/
(with high probability)
{\em a set that contains more than $m$ strings
that are $\e/2$-H-far apart from one another}.
In this case, with high probability, the projection
of the $s$ samples on $\ell=O(\e^{-1}\log m)$ random locations
would yield more than $m$ different $\ell$-bit strings.
Hence, given oracle access to the samples $x^{(1)},...,x^{(s)}$,
our tester proceeds as follows.

\smallskip
\begin{enumerate} 
\item Selects uniformly an $\ell$-subset $J$ of $[n]$.
\item For every $i\in[s]$, queries $x^{(i)}$ at the locations $j\in J$,
obtaining $y^{(i)}=x^{(i)}_J$, where $x^{(i)}_J$ denotes
the restriction of $x^{(i)}$ to the locations in $J$.
\item Accepts if and only if $|\{y^{(i)}:i\in[s]\}|\leq m$.
\end{enumerate} 
Clearly if $X$ has support size at most $m$,
then the foregoing tester always accepts.
On the other hand, as shown next,
if $X$ is $\e$-far from having support size at most $m$,
then, with very high probability over the choice of the $s$ samples,
there exists a set $I\subset[s]$ of size $m+1$ such that
for every $i_i\neq i_2\in I$ the relative Hamming distance
between $x^{(i_1)}$ and $x^{(i_2)}$ is at least $\e/2$.

To prove the foregoing claim,
we consider a mental experiment in which we try to find $m+1$
strings that are pairwise far apart in $m+1$ sampling phases.
For $i\in[m]$, suppose that $w^{(1)},\dots,w^{(i)}$
are pairwise $\e/2$-H-far apart strings
that were found in the first $i$ phases.
Then, assuming that $X$ is $\e$-far from having support size at most $m$,
it must be that at least $\e/2$ of the probability weight of $X$
resides on strings that are $\e/2$-H-far from each of the $w^{(j)}$'s
(for all $j\in[i]$).
Hence, a sampled string taken from $X$ is $\e/2$-H-far from all
these $w^{(j)}$'s with probability at least $\e/2$.
Once such a sampled string is found, we denote it $w^{(i+1)}$,
and proceed to the next phase.
It follows that, with probability $1-\exp(-\Omega(m))$,
the process is completed successfully (i.e., $m+1$ strings are found)
while using $O(m/\e)$ samples.

\paragraph{Testing equality of distributions.}
Our proof of Theorem~\ref{equal-distributions:ithm} relies on
the following claim, which may be of independent interest.

\BCM{\em(Typically, the distance between $X$ and $Y$
is preserved by the distance between~$X_J$ and~$Y_J$
(see Claim~\ref{two-distributions:clm})):}
\label{two-distributions:project-clm}
Suppose that $X$ is $\e$-far from $Y$,
and that each distribution has support size at most $m$.
Then, with probability $1-o(1)$ over the choice
of $J\in{{[n]}\choose{O(\e^{-1}\log m)}}$,
it holds that $X_J$ is $0.3\e$-far from $Y_J$,
\ECM
The constant $0.3$ can be replaced by any constant smaller than~1,
whereas the dependence on $\log m$ is essential
(e.g., consider $X$ and $Y$ that are each uniformly distributed
on a randomly selected $m$-subset of $\bitset^n$, where $m=2^{o(n)}$).

\subsubsection{Lower bounds}
Several of our lower bounds are obtained by transporting
lower bounds from the standard distribution-testing model.
Typically, we transform distributions over an alphabet $\Sigma$
to distributions over $\bitset^n$ by using an error correcting
code $C:\Sigma\to\bitset^n$ that has constant relative distance
(i.e., $\Delta_H(C(\sigma),C(\tau))=\Omega(1)$
for every $\sigma\neq\tau\in\Sigma$).

For example, when proving a lower bound on testing the support size,
we transform a random variable $Z$ that ranges over $\Sigma$
to the random variable $C(Z)$.
Clearly, if $Z$ has support size at most $m$, then so does $C(Z)$.
On the other hand,
if $Z$ is TV-far from having a support of size at most $m$,
then $C(Z)$ is far (under Definition~\ref{dist:def}) from
being supported on (at most) $m$ {\em codewords}.
However, we need to show that $C(Z)$ is far from being supported
on {\em any}\/ (subset of at most)~$m$ strings.
Specifically, we need to show that transporting
a probability mass from more than~$m$ codewords
to at most $m$ strings requires a constant fraction of this mass
to be transported over a constant (relative Hamming) distance.
Intuitively, this is the case because transporting a probability
mass to the vicinity of a codeword does not reduce the support size
unless a similar (or larger) mass is transported from a different codeword.


\subsection{Related work}
Our work is informed by the vast literature on distribution testing
(see~\cite[Chap.~11]{G:pt} and~\cite{C:survey})
and on property testing at large (see~\cite{G:pt}).
Here we only mention a prior model that refers to
partial information obtained from individual samples,
and a subsequent work that studies the \DHOm.

\subsubsection{A different model of partial information from samples}
A sequence of works, initiated by~\cite{ACT},
studies a model in which one may obtain only partial information
regarding each of the samples drawn from the distribution.
Specifically, the tester (or learner) may obtain only $\ell\geq1$
``bits of information'' from each of the samples; however,
unlike in the \DHOm, these $\ell$ bits may be
an arbitrary function of the entire sample
(rather than actual bits in the representation of the sample),
which is actually viewed as an atom,
and the distance between distributions is the total variation distance
(rather than Definition~\ref{dist:def}).

\subsubsection{Subsequent work}
Following the initial posting of our work~\cite{GR:tr},
a study of {\em index invariant}\/ properties in the \DHOm,
was initiated in~\cite{CFGMS},
where a property $\D$ (of distributions over $n$-bit strings)
is called {\sf index invariant}
if a distribution $X=X_1\cdots X_n$ is in $\D$
if and only if, for every permutation $\phi:[n]\to[n]$
(on the indices of strings in $\bitset^n$),
it holds that $X_{\phi(1)}\cdots X_{\phi(n)}$ is in $\D$.
Note that the class of index-invariant properties extends the
class of label-invariant properties, since the invariance
condition is much weaker: It only refers to relabeling
of strings by rearranging their bits
(i.e., it only considers permutations $\pi:\bitset^n\to\bitset^n$
that satisfy $\pi(x_1\cdots x_n)=x_{\phi(1)}\cdots x_{\phi(n)}$
for some permutation $\phi:[n]\to[n]$).

One notable result of~\cite{CFGMS} is that such properties
can be tested within query complexity that is upper-bounded
by a function of the VC-dimension (and the proximity parameter).
They also show that their complexity bounds are tight,
that index-invariance is essential for these bounds,
and that there exists a (tight) quadratic gap between
adaptive and non-adaptive testers of such properties.

\subsection{Organization}
We start, in Section~\ref{support+uniform+grained:sec},
with results that refer to a few natural properties
of distributions that were studied previously
in the context of the standard distribution-testing model.
Specifically, we present the testers asserted in
Theorems~\ref{closed-mapping:ithm} and~\ref{support+uniform:ithm},
and establish the lower bounds
claimed in Theorem~\ref{lb:support+uniform:ithm}.

Next, we turn to the general result captured by Theorem~\ref{tentative:ithm},
and present its proof in Section~\ref{selt-correction:sec}.
In Section~\ref{material-ideal:sec} we study several types
of distributions that arise naturally in the context of the \DHOm;
that is, we consider distributions that capture
random variations of some ideal objects.
Lastly, in Section~\ref{tuples:sec}, we extend our treatment
to testing tuples of distributions, and present a tester
for the set of pairs of identical distributions
(proving Theorem~\ref{equal-distributions:ithm}).

\section{Support Size, Uniformity, and Being Grained}
\label{support+uniform+grained:sec}
In this section we consider three natural
label-invariant properties (of distributions).
These properties refer to the support size,
being uniform (over some subset),
and being $m$-grained (i.e., each string appears
with probability that is an integer multiple of $1/m$).
Recall that $\D$ is a {\sf label-invariant property} of distributions
over $\bitset^n$ if for every bijection $\pi:\bitset^n\to\bitset^n$
and every distribution $X$, it holds that $X$ is in $\D$
if and only if $\pi(X)$ is in $\D$.
Label-invariant properties of distributions are of general interest
and are particularly natural in the \DHOm, in which we wish to avoid
reading samples in full. In this section we explore the possibility
of obtaining testers for such properties.

We first present testers for these properties (in the \DHOm),
and later discuss related ``triviality results'' and lower bounds.
Our testers (for the \DHOm) are derived by emulating
testers for the standard (distribution testing) model.
The lower bounds justify this choice retroactively.

\subsection{Testers}
Our (\DHO-model) testers for support size,
being uniform (over some subset), and being $m$-grained
are obtained from a general result that refers to
arbitrary properties (of distributions) that satisfy
the following condition.

\BD{\rm(Closure under mapping):}
We say that a property $\D$ of distributions over $n$-bit strings
is {\sf closed under mapping}
if for every $f:\bitset^n\to\bitset^n$
it holds that if $X$ is in $\D$ then $f(X)$ is in $\D$.
\ED
Note that closure under mapping implies being label-invariant
(i.e., for every bijection $\pi:\bitset^n\to\bitset^n$,
consider both the mapping $\pi$ and $\pi^{-1}$).
Indeed, the properties of having support size at most $m$
and being $m$-grained are closed under mapping,
but uniformity is not.

\BT{\em(Testing distributions that are closed under mapping):}
\label{closed-mapping:thm}
Suppose that $\D=\{\D_n\}$ is testable
with sample complexity $s(n,\e)$ in the standard model,
and that each $\D_n$ is closed under mapping.
Then, $\D$ is testable in the \DHOm\
with query complexity $\tildeO(\e^{-1}\cdot s(n,\e/2))$.
Furthermore, the resulting tester uses $3\cdot s(n,\e/2)$ samples,
makes {\rm(the same)} $O(\e^{-1}\log(s(n,\e/2)/\e))$
uniformly distributed queries to each sample,
and preserves one-sided error of the original tester.
\ET
The factor of~3 in the sample complexity is due to modest error reduction
that is used to compensate for the small error that is introduced by
our main strategy.
Recall that a tester of sample complexity $s$
in the standard distribution-testing model constitutes
a tester of sample complexity $s$ in the \DHOm,
alas this tester has query complexity $n\cdot s$.
\medskip

\BPF
The key observation is that, since $\D$ is closed under mapping,
for any $\ell$-subset $J\subseteq[n]$, it holds that if $X$ is in $\D$,
then $X_J0^{n-\ell}$ is in $\D$, whereas we can test $X_J0^{n-\ell}$
for membership in~$\D$ with $\ell$ queries per sample.
Furthermore, as shown below, if $X$ is $\e$-far from $\D$,
then, for a random set $J\subseteq[n]$ of size $\ell=O(\e^{-1}\log(s/\e))$,
the original tester would have rejected $X_J0^{n-\ell}$,
when invoked with proximity parameter $\e/2$.
The foregoing claim relies on the fact that we may assume,
without loss of generality,
that the original tester (which operates in the standard model)
rules according to the collision pattern among the samples that it gets.
Hence, it suffices to show that the collision pattern among $s$ samples
of $X_J$ is statistically close to the collision pattern of $s$ samples
drawn from a distribution $Y$ that is $\e/2$-far from $\D$.
Let us first spell out the proposed tester.

Our starting point is the guaranteed tester $T$,
which operates in the standard distribution-testing model.
Hence, we may assume, without loss of generality,
that $T$ is {\em label-invariant}\/ (see, e.g.,~\cite[Thm.~11.12]{G:pt}),
which means that it rules according to the collision pattern
that it sees among its samples
(i.e., the number of $t$-way collisions for each $t\geq2$).
In particular, if the collision pattern of $s$ samples of $X_J$
(equiv., of $X_J0^{n-\ell}$)
is statistically close to the collision pattern of $s$ samples of $Y$
that is $\e/2$-far from $\D$, then $T$ will reject $X_J0^{n-\ell}$.
(Indeed, in the analysis, we shall present such an $Y$.)

\myparagraph{The actual tester.}
Let $T$ be the guaranteed tester of sample complexity $s:\N\times[0,1]\to\N$.
Using $T$, on input parameters $n$ and $\e$,
when given $s=s(n,\e/2)$ samples, denoted $x^{(1)},....,x^{(s)}$,
that are drawn independently from a tested distribution $X$,
we proceed as follows.
\begin{enumerate}
\item We select a set $J\subseteq[n]$ of size $\ell=O(\e^{-1}\log(s/\e))$
uniformly at random and query each of the samples at each location in $J$.
Hence, we obtain $x_J^{(1)},...,x_J^{(s)}$.

(Recall that $x_J^{(i)}$ denotes the restriction of $x^{(i)}$ to $J$.)
\item Invoking $T$ with proximity parameter $\e/2$,
we output $T'(x_J^{(1)},...,x_J^{(s)})$,
where
\begin{equation}\label{T':eqdef}
T'(z^{(1)},...,z^{(s)}) = T(n,\e/2;z^{(1)}0^{n-\ell},...,z^{(s)}0^{n-\ell}).
\end{equation}
That is, we invoke $T$ on $s$ samples of the distribution $X_J0^{n-\ell}$,
where these $s$ samples are obtained by padding
the strings $x_J^{(1)},...,x_J^{(s)}$ obtained in Step~1.
\end{enumerate}
As observed upfront, if $X$ is in $\D$, then so is $X_J0^{n-\ell}$,
for any choice of $J$.
Hence, our tester accepts each distribution in $\D$
with probability that is lower-bounded by the corresponding
lower bound of~$T$. In particular, if $T$ has one-sided error,
then so does our tester.

We now turn to the analysis of the case that $X$ is $\e$-far from $\D$.
In this case, we proceed with a mental experiment in
which we define, for each choice of $J$,
a random variable $Y=Y(J)$ such that
for most $J$'s it holds that
(i)~$Y_J\equiv X_J$, (ii)~$Y$ is $\e/2$-close to $X$,
and (iii)~the collision pattern of $s$ samples of $Y_J$
is statistically close to the collision pattern of $s$ samples of $Y$.
Note that Condition~(ii) implies that $Y$ is $\e/2$-far from $\D$,
which means that $T$ should reject $s$ samples of $Y$ (whp),
Condition~(iii) implies that $T$ should also reject $s$
samples of $Y_J0^{n-\ell}$ (whp), whereas Condition~(i)
implies that the same holds for samples of $X_J0^{n-\ell}$,
which in turn means that our tester rejects $X$ (whp).
In order to materialize the foregoing plan, we need a few definitions.

\myparagraph{Initial definitions and observations.}
For integers $\ell\leq n$ and $s$,
and a generic random variable~$X$ that ranges over $\bitset^n$,
we consider a sufficiently large $s'=O(s^2\cdot\ell)$,
and use the following definitions.
\BI
\item For any fixed $\ell$-subset $J$,
we say that $\sigma\in\bitset^\ell$ is {\sf $J$-heavy (w.r.t $X$)}
if $\prob[X_J\!=\!\sigma]\geq{0.01}/{s^2}$.

(The definition of $J$-heavy is aimed at guaranteeing that
the probability that a collision on a non-$J$-heavy string
occurs among the $J$-restrictions of $s$ samples of $X$ is small;
that is, denoting the set of non-$J$-heavy strings by $L$,
it holds that
$\prob_{x^{(1)},...,x^{(s)}\sim X}
 [\exists i\neq j\rst x^{(i)}_J\!=\!x^{(j)}_J\!\in\!L]<0.005$.)

\item For any fixed $\ell$-subset $J$,
we say that a sequence
of $s'$ strings $(w^{(1)},....,w^{(s')})\in(\bitset^n)^{s'}$
is {\sf $J$-good (for $X$)}
if its $J$-restrictions hit all $J$-heavy strings;
that is, for every $J$-heavy string~$\sigma$
there exists $i\in[s']$ such that $w^{(i)}_J=\sigma$.

Note that, for every $J$,
$$\prob_{w^{(1)},....,w^{(s')}\sim X}
     [\mbox{\rm $(w^{(1)},....,w^{(s')})$ is $J$-good}] = 1-o(1),$$
because the probability that some $J$-heavy string is not
hit by any $w^{(i)}_J$ is upper-bounded by
$2^{\ell}\cdot(1-0.01/s^2)^{s'} = 2^{\ell}\cdot\exp(-\Omega(s'/100s^2)) = o(1)$.

(Here we used the fact that $s'=\Omega(s^2\cdot\ell)$.)
\item We say that $(w^{(1)},....,w^{(s')})$ is {\sf good (for $X$)}
if it is $J$-good for a $1-o(1)$ fraction of the $\ell$-subsets $J$'s.

By an averaging argument,
$$\prob_{w^{(1)},....,w^{(s')}\sim X}
     [\mbox{\rm $(w^{(1)},....,w^{(s')})$ is good}] = 1-o(1).$$
Actually, we shall only use the fact that there exists
a good sequence of $w^{(i)}$'s.
\EI
We fix an arbitrary good (for $X$) sequence $(w^{(1)},....,w^{(s')})$
for the rest of the proof.

\myparagraph{The definition of $Y$.}
Recall that,
with probability $1-o(1)$ over the choice of $J\in{{[n]}\choose\ell}$,
it holds that $(w^{(1)},...,w^{(s')})$ is $J$-good (for $X$),
which means that {\em all $J$-heavy strings}\/ (w.r.t $X$)
{\em appear among the $J$-restrictions of the $w^{(i)}$'s}.
Fixing such a (typical) set $J$,
let $I=I(J)$ be a maximal set of indices $i\in[s']$
such that the $w^{(i)}_J$'s are distinct;
that is, $R=R(J)\eqdef\{w^{(i)}_J:i\!\in\!I\}$ has size $|I|$
and equals $\{w^{(i)}_J:i\!\in\![s']\}$.
We stress that $R$ contains all $J$-heavy strings (w.r.t $X$),
which means that for every $\sigma\not\in R$
it holds that $\prob[X_J\!=\!\sigma]<0.01/s^2$,
and that each $J$-heavy string corresponds to a unique $i\in I$.
We now define $Y$ by selecting $x\sim X$, and outputting $w^{(i)}$
if $x_J=w^{(i)}_J$ for some $i\in I$,
and outputting $x$ itself otherwise (i.e., if $x_J\not\in R(J)$);
that is,
\begin{equation}
\label{Y-vs-X:eq}
\prob[Y\!=\!x] = \left\{
\begin{array}{lll}
\prob[X_J\!=\!w^{(i)}_J] & & \mbox{\rm if $x=w^{(i)}$ for $i\in I$} \\
0 & & \mbox{\rm if $x_J\in\{w^{(i)}_J:i\!\in\!I\}$
               but $x\not\in\{w^{(i)}:i\!\in\!I\}$} \\
\prob[X\!=\!x] & & \mbox{\rm if $x_J\not\in\{w^{(i)}_J:i\!\in\!I\}$}
\end{array} \right.
\end{equation}
Note that $Y_J\equiv X_J$, which establishes Condition~(i) of our plan.
Turning to Condition~(ii), we now prove that, for a typical $J$,
it holds that $Y$ is $\e/2$-close to $X$.

\Bcm{\em(Typically, $Y$ is $\e/2$-close to $X$):}
\label{closed-mapping:clm1}
With probability $1-o(1)$ over the choice of $J$,
the corresponding $Y=Y(J)$ is $\e/2$-close to $X$.
\Ecm

\Bpf
The key observation is that $Y$ differs from $X$ only
when $X_J\in\{w^{(i)}_J:i\!\in\!I(J)\}=\{w^{(i)}_J:i\!\in\![s']\}$.
In this case, strings that are $\e/4$-H-close to $\{w^{(i)}:i\!\in\!I(J)\}$
contribute at most $\e/4$ units
(to the distance between $X$ and $Y$ (as in Definition~\ref{dist:def})),
and so we upper-bound the probability mass of strings $x\sim X$
that are $\e/4$-H-far from $\{w^{(i)}:i\!\in\!I(J)\}$
but satisfy $x_J\in\{w^{(i)}_J:i\!\in\![s']\}=R(J)$.
Actually, letting $W\eqdef\{w^{(i)}:i\!\in\![s']\}$
and $F_{\e'}(S)$ denote the set
of $n$-bit strings that are $\e'$-H-far from $S$,
for every $x\in F_{\e/4}(W)$,
we let $\mbox{\tt Bad}_x(J)$ denote the event $x_J\in R(J)$.
Letting $\mbox{\tt Bad}_x(J)=0$ for any $x$ that is $\e/4$-H-close to $W$,
we have
\begin{eqnarray*}
\prob_{J,X}[\mbox{\tt Bad}_X(J)]
&=& \Exp_{x\sim X}\left[\prob_{J\in{{[n]}\choose\ell}}
	[\mbox{\tt Bad}_x(J)]\right] \\
&\leq& \max_{x\in F_{\e/4}(W)}
	\left\{\prob_{J\in{{[n]}\choose\ell}}[\mbox{\tt Bad}_x(J)]\right\} \\
&\leq& \max_{x\in F_{\e/4}(W)}
	\left\{\sum_{i\in[s']}
		\prob_{J\in{{[n]}\choose\ell}}[x_J=w^{(i)}_J] \right\} \\
&\leq& \sum_{~~i\in[s']~~}\max_{x\in F_{\e/4}(\{w^{(i)}\})}
	\left\{\prob_{J\in{{[n]}\choose\ell}}[x_J=w^{(i)}_J] \right\} \\
&<& s'\cdot(1-(\e/4))^{\ell},
\end{eqnarray*}
which is $o(\e)$ by the definition of $\ell=O(\e^{-1}\log(s/\e))$
(and $s'=\tildeO(s^2/\e)$, where we actually use $s'\leq\poly(s/\e)$).
Hence, with probability $1-o(1)$ over the choice of $J$,
it holds that the probability that $x\sim X$
is $\e/4$-H-far from $W=\{w^{(i)}:i\!\in\![s']\}$
but satisfies $x_J\in R(J)=\{w^{(i)}_J:i\!\in\![s']\}$ is at most $\e/4$;
that is,
\begin{equation}
\label{typical-J:eq}
\prob_{J\in{{[n]}\choose\ell}}
	\Big[\prob_{x\sim X}\left[
	   \mbox{\rm $x\in F_{\e/4}(W)$ but $x_J\!\in\!R(J)$} \right]
   \leq\e/4 \Big]
\;=\;1-o(1).
\end{equation}
It follows that, with probability $1-o(1)$ over the choice of $J$,
it holds that $Y$ is $\e/2$-close to $X$,
where one term of $\e/4$ is due to the $x$'s that are $\e/4$-H-close
to $\{w^{(i)}:i\!\in\![s']\}$
and the other term is due to the probability mass of $x$'s
that are $\e/4$-H-far to $\{w^{(i)}:i\!\in\![s']\}$
but satisfies $x_J\in R(J)$.
\Epf

\medskip
Recalling that $X$ is $\e$-far from~$\D$, for a typical $J$,
Claim~\ref{closed-mapping:clm1} implies that $Y$ is $\e/2$-far from~$\D$,
which implies that $Y$ is $\e/2$-TV-far from $\D$,
which implies that (with probability at least $2/3$)
the tester $T$ rejects $Y$
(i.e., rejects when fed with $s$ samples selected according to $Y$).
However, we are interested in the probability that
our tester (rather than $T$) rejects $X$ (rather than $Y$).

\Bcm{\em(Typically, our tester rejects $X$):}
\label{closed-mapping:clm2}
Suppose that $(w^{(1)},...,w^{(s')})$ is $J$-good for $X$
and that the corresponding $Y=Y(J)$ is $\e/2$-far from $\D$.
Then, our tester rejects $X$ with probability at least $0.66$.
\Ecm

\Bpf
Recalling that $Y_J\equiv X_J$,
while relying on the hypothesis
that ${\ov w}\eqdef(w^{(1)},...,w^{(s')})$ is $J$-good (for $X$),
we observe that the probability that our tester rejects $X$ equals
\[\begin{array}{llr}
& \prob_{x^{(1)},...,x^{(s)}\sim X}
                [T'(x_J^{(1)},....,x_J^{(s)})\!=\!0] & \\
& = \prob_{y^{(1)},...,y^{(s)}\sim Y}
               [T'(y_J^{(1)},....,y_J^{(s)})\!=\!0]
& [\mbox{\rm using $Y_J\equiv X_J$}] \\
& = \prob_{y^{(1)},...,y^{(s)}\sim Y}
             [T(n,\e/2;y_J^{(1)}0^{n-\ell},....,y_J^{(s)}0^{n-\ell})\!=\!0]
& [\mbox{\rm definition of $T'$}] \\
& = \prob_{y^{(1)},...,y^{(s)}\sim Y}
             [T(n,\e/2;y^{(1)},....,y^{(s)})\!=\!0]
     \pm\frac{{s\choose2}}{100\cdot s^2}
& ~~~~~~~~~[\mbox{\rm see next}]
\end{array}\]
where the last (approximate) equality is justified as follows
(based on the definition of $Y$
and the hypothesis that ${\ov w}$ is $J$-good (for $X$)).
\BI
\item On the one hand, the equality-relations between samples of $Y$
with a $J$-restriction in $R$ are identical to those of their $J$-restrictions
(i.e., for $y^{(i)}_J,y^{(j)}_J\in R$,
it holds that $y^{(i)}=y^{(j)}$ iff $y^{(i)}_J=y^{(j)}_J$).
This holds because (by the definition of $Y$), for each $\sigma\in R$,
there is a unique $y$ in the support of $Y$ such that $y_J=\sigma$
(i.e., $y=w^{(i)}$ such that $w^{(i)}_J=\sigma$).
\item On the other hand, the probability of collision among
the $J$-restrictions of the other samples
(i.e., those with a $J$-restriction in $\bitset^\ell\setminus R$)
is upper-bounded by ${s\choose2}\cdot\frac{1}{100\cdot s^2}<0.005$.
This is because (by the hypothesis that ${\ov w}$ is $J$-good)
these $J$-restrictions are all non-heavy
	(and $\sum_{\sigma\in L} p_\sigma^2\leq\max_{\sigma\in L}\{p_\sigma\}$,
for non-negative $p_\sigma$'s that sum-up to at most~1).
Needless to say, the collision probability between
these (other) samples themselves can only be smaller.
\EI
Indeed, the foregoing analysis establishes Condition~(iii) of our plan
(i.e., the collision pattern of~$s$ samples of $Y_J$ is $0.005$-close
to the collision pattern of $s$ samples of $Y$).

\smallskip
\noindent
It follows that our tester rejects $X$ with probability
at least $\frac23-0.005>0.66$,
where the first term lower-bounds the probability that $T$ rejects
when presented with $s$ samples of $Y$.
\Epf

\myparagraph{Conclusion.}
Using the hypothesis that $(w^{(1)},...,w^{(s')})$ is good (for $X$),
with probability $1-o(1)$ over the choice of $J\in{{[n]}\choose\ell}$,
it holds that $(w^{(1)},...,w^{(s')})$ is $J$-good (for $X$)
and (by Claim~\ref{closed-mapping:clm1})
the corresponding $Y=Y(J)$ is $\e/2$-close to $X$.
Assuming that $X$ is $\e$-far from $\D$,
using both conclusions,
it follows (by Claim~\ref{closed-mapping:clm2})
that our tester rejects $X$ with probability at least $0.66-o(1)$.
Using mild error reduction (via three experiments), the theorem follows.
\EPF

\BCR{\em(Testers for support size, uniformity, and $m$-grained in the \DHOm):}
\label{support+uniform:cor}
For any~$m$, the following properties of distributions over $\bitset^n$
can be tested in the \DHO\ model using $\poly(1/\e)\cdot\tildeO(m)$ queries:

\begin{enumerate}
\item The set of all distributions having support size at most $m$.

Furthermore, the tester uses $O(m/\e)$ samples,
makes $O(\e^{-1}\log(m/\e))$ queries to each sample,
and has one-sided error probability.

\item The set of all distributions that are uniform over some set of size $m$.

Furthermore, the tester uses $O(\e^{-2}m\log m)$ samples,
and makes $q=O(\e^{-1}\log(m/\e))$ queries to each sample.

\item The set of all distributions that are $m$-grained.

Furthermore, the tester uses $O(\e^{-2}m\log m)$ samples,
and makes $O(\e^{-1}\log(m/\e))$ queries to each sample.
\end{enumerate}
Moreover, all testers make the same uniformly distributed queries
to each of their samples.
\ECR

\BPF
For Parts~1 and~3 we present testers for the standard model
and apply Theorem~\ref{closed-mapping:thm},
whereas for Part~2 we observe that a (\DHOm) tester
for $m$-grained distributions will do.

Let us start with Part~2.
The key observation is that any distribution
that is uniform over some $m$-subset is $m$-grained,
whereas any distribution that is $m$-grained
is $\frac{\ceil{\log_2 m}}{n}$-close (under Definition~\ref{dist:def})
to being uniform over some set of $m$ elements
(e.g., by modifying the first $\ceil{\log_2m}$ bits
in each string in the support).%
\footnote{Saying that $X$ is $m$-grained means that it is uniform
on a multi-set $\{x^{(1)},\dots,x^{(m)}\}$ of $n$-bit strings.
We modify $X$ by replacing each $x^{(i)}$ by $y^{(i)}$
such that $y^{(i)}$ encodes the binary expansion of $i-1$ in the
first $\ell=\ceil{\log_2m}$ locations and equals $x^{(i)}$ otherwise.
That is, we set $y^{(i)}_{j}$ to equal the $j^\xth$ bit
in the binary expansion of $i-1$ if $j\in[\ell]$,
and $y^{(i)}_j=x^{(i)}_j$ otherwise (i.e., if $j\in \{\ell+1,...,n\}$).}
Hence, for $\e>2\cdot\frac{\ceil{\log_2m}}{n}$,
we test uniformity over $m$-subsets by testing for
being $m$-grained (using proximity parameter $\e/2$),
while noting that being $\e$-far from uniformity over an $m$-subset
implies being $(\e/2)$-far from $m$-grained.
Lastly, if $\e\leq\frac{2\ceil{\log_2m}}{n}$,
then we can afford reading entirely each sample,
since $n=O(\e^{-1}\log m)$.

Turning to Parts~1 and~3,
it is tempting to use known (standard model) testers
of complexity $O(\e^{-2}m/\log m)$ for these properties (cf.~\cite{VV-jacm}),
while relying on the fact that these properties are label-invariant.
However, these bounds hold only when the tested distribution ranges over
a domain of size $O(m)$, and so some additional argument is required.
Furthermore, this may not allow us to argue that
the tester for support-size has one-sided error.
Instead, we present direct (standard model) testers
of sample complexity $O(m/\e)$ and $\tildeO(m/\e^2)$, respectively.

\mypar{Testing support size.}
On input parameters $n$ and $\e$,
given $s=O(m/\e)$ samples, denoted $x^{(1)},....,x^{(s)}$,
that are drawn independently from a tested distribution $X$,
we accept if and only if $|\{x^{(i)}:i\!\in\![s]\|\leq m$.
Suppose that $X$ is $\e$-TV-far from having support size at most $m$,
and note that for any set $S$ of at most $m$ strings
it holds that $\prob[X\!\not\in\!S]>\e$.
Then, for each $t\in[s-1]$, either $W_t=\{x^{(i)}:i\!\in\![t]\}$
has size exceeding $m$ or $\prob[x^{(t+1)}\!\not\in\!W_t]>\e$.
It follows that $\prob[|W_s|\leq m]<(1-\e)^s=\exp(-\Omega(m))$.

\mypar{Testing the set of $m$-grained distributions.}
On input parameters $n$ and $\e$, we set $s=O(m\log m)$
and $s'=O(\e^{-2}m\log m)$.
Given $s+s'$ samples, denoted $x^{(1)},....,x^{(s+s')}$,
that are drawn independently from a tested distribution $X$,
we proceed in two steps.

\begin{enumerate} 
\item We construct $W=\{w^{(i)}:i\!\in\![s]\}$,
the set of strings seen in the first $s$ samples.

(We may reject if $|W|>m$, but this is inessential.)
\item For each $w\in W$, we approximate $\prob[X\!=\!w]$
by $p_w\eqdef|\{i\!\in\![s']:x^{(s+i)}\!=\!w\}|/s'$.
We reject if we either encountered a sample not in $W$
or one of the $p_w$'s is not within a $1\pm0.1\e$ factor
of a positive integer multiple of $1/m$.
\end{enumerate} 
Note that if $X$ is $m$-grained, then, with high probability, $W$
equals the support of $X$, and (whp) each of the $p_w$'s
is within a $1\pm0.1\e$ factor of a positive integer multiple of $1/m$.
On the other hand, suppose that $X$ is accepted with high probability.
Then, for any choice of $W$ (as determined in Step~1),
for each $w\in W$, it holds that $\prob[X\!=\!w]=(1\pm0.1\e)\cdot p_w$,
since we may assume that $\prob[X\!=\!w] \geq 1/2m$
(as otherwise $p_w\geq(1-0.1\e)/m$ is unlikely).
Furthermore, $\prob[X\!\not\in\!W]<0.1\e$.
It follows that $X$ is $\e$-TV-close to being $m$-grained.
\EPF

\subsection{Triviality results}
\label{triviality:sec}
An obvious case in which testing is trivial
is the property of all distributions (on $n$-bit strings)
that have support size $2^n$.
In this case, each distribution is infinitesimally close
(under Definition~\ref{dist:def})
to being supported on all $2^n$ strings.
A less obvious result is stated next.

\begin{observation}[Triviality of testing $2^n$-grained distributions in the \DHOm]
\label{triviality:obs}
Under Definition~\ref{dist:def}, every distribution over $\bitset^n$
is $O(\frac{\log n}{n})$-close to being $2^n$-grained.
Furthermore, for every $\ell'\in[O(1)+\log_2\log_2n]$,
every distribution over $\bitset^n$
is $O(\frac{\log n}{n})$-close to being $2^{n-\ell'}$-grained.
\end{observation}
In contrast, in the standard (distribution testing) model,
testing whether a distribution (over $\bitset^n$)
is $2^{n-O(1)}$-grained requires $2^{0.99\cdot n}$ samples~\cite{GR:gr}.
\medskip

\BPF
We first show that, for every $\ell\in\N$, it holds that
every distribution over $\bitset^n$ is $\frac{\ell}{n}$-close
to a distribution that is supported by $\bitset^{n-\ell}0^\ell$.
Next, we show that each distribution of the latter type
is $2^{-\ell}$-close to being $2^n$-grained.
Letting $\ell=\floor{\log_2n}$, the main claim follows.

In the first step, given an arbitrary distribution $X$,
we consider the distribution $X'$ obtained by setting the
last $\ell$ bits of $X$ to zero; that is, let
$\prob[X'\!=\!x'0^\ell]=\sum_{x''\in\bitset^\ell}\prob[X\!=\!x'x'']$.
Then, $X'$ is $(\ell/n)$-close to $X$
(according to Definition~\ref{dist:def}).

In the second step, using $X'$, we consider $X''$ obtained
by letting $\prob[X''\!=\!x'0^\ell]$
equal $2^{-n}\cdot\floor{2^n\cdot\allowbreak\prob[X'\!=\!x'0^\ell]}$,
and assigning the residual probability to (say)~$1^n$.
Then, $X''$ is $2^n$-grained and is at total variation
distance at most $2^{n-\ell}\cdot2^{-n}=2^{-\ell}$ from $X'$,
since the support size of $X'$ is at most $2^{n-\ell}$.
Hence, $X''$ is $(\frac{\ell}{n}+2^{-\ell})$-close to $X$.

The furthermore claim follows by redefining $X''$
such that $\prob[X''\!=\!x'0^\ell]$
equal $2^{-(n-\ell')}\cdot\floor{2^{n-\ell'}\cdot\prob[X'\!=\!x'0^\ell]}$.
In this case $X''$ is $2^{n-\ell'}$-grained and is at total variation
distance at most $2^{n-\ell}\cdot2^{-(n-\ell')}=2^{-(\ell-\ell')}$ from $X'$,
which means that it is $O(\ell/n)$-close to $X$,
since $\ell'=O(\log\ell)$.
\EPF

\paragraph{Non-triviality results.}
It is easy to see that any property of distributions that includes
only distributions having a support of size $2^{n-\Omega(n)}$
is non-trivial in the sense that {\em not}\/ all distributions
are close to it under Definition~\ref{dist:def}.
This is the case because any such distribution
is far from the uniform distribution over $\bitset^n$
(since, w.h.p., a uniformly distributed $n$-bit string
is at Hamming distance $\Omega(n)$ from a set
that contains $2^{n-\Omega(n)}$ strings).
Additional non-triviality results follow from the lower bounds
presented in Section~\ref{lower-bounds:sec}.

\subsection{Lower bounds}
\label{lower-bounds:sec}
We first consider three notions of uniformity:
Uniformity over the entire potential support (i.e., all $n$-bit strings),
uniformity over the support of the distribution
(where the size of the support is not specified),
and uniformity over a support of a specified size
(also called ``parameterized uniformity'').
In all three cases (as well as in the results regarding testing
support size and the set of grained distributions),
we prove lower bounds on the sample (and query) complexity of testing
the corresponding property in the \DHOm.

As usual, the lower bounds refer to testing with $\e=\Omega(1)$; that is,
to the case that the proximity parameter is set to some positive constant.
Our proofs rely on the standard methodology by which
a lower bound of $L$ on the complexity of testing
is proved by presenting two distributions $X$ and $Y$
that an algorithm of complexity $L-1$ cannot distinguish
(with constant positive gap)%
\footnote{We say that $A$ {\sf distinguishes} $s$ samples of $X$ from $s$
samples of $Y$ {\sf with gap $\gamma$} if
$$\left|\prob_{z_1,\dots,z_s\sim X}[A(z_1,\dots,z_s)\!=\!1]
 -\prob_{z_1,\dots,z_s\sim Y}[A(z_1,\dots,z_s)\!=\!1]\right|
\geq \gamma.$$\label{gap:fn}}
such that $X$ has the property
and $Y$ is $\Omega(1)$-far from having the property
(cf.~\cite[Thm.~7.2]{G:pt}).
In fact, typically, at least one of the two distributions
will be claimed to exist using a probabilistic argument;
that is, we shall actually prove that there exists two
distributions $X_0$ and $Y_0$ (over $\bitset^n$)
such that, for a random bijection $\pi:\bitset^n\to\bitset^n$,
setting $X=\pi(X_0)$ and $Y=\pi(Y_0)$ will do.

\BOB{\em(Lower bound on testing uniformity over $\bitset^n$):}
\label{uniform-on-all:obs}
For every $c\in(0,0.5)$ there exists $\e>0$ such that 
testing with proximity parameter $\e$ whether
a distribution is uniform over $\bitset^n$
requires $2^{c\cdot n}$ samples in the \DHOm.
\EOB

\BPF
Let $S$ be a random $2^{2c\cdot n}$-subset of $\bitset^n$,
and $X$ be uniform over $S$. Then, a sample of $s=o(2^{c n})$
strings does not allow for distinguishing
between $X$ and the uniform distribution over $\bitset^n$; that is,
for every decision procedure $D:(\bitset^n)^s\to\bitset$,
and for all but a $o(1)$ fraction of the $2^{2c\cdot n}$-subsets $S$
it holds that
$$\prob_{x^{(1)},\dots,x^{(s)}\in S}[D(x^{(1)},\dots,x^{(s)})\!=\!1]
 = \prob_{x^{(1)},\dots,x^{(s)}\in\bitset^n}[D(x^{(1)},\dots,x^{(s)})\!=\!1]
  \pm o(1).$$
(Intuitively, this is the case because $s$ random samples from
a random set $S$ are distributed almost identically to $s$
random samples from the uniform distribution over $n$-bit strings.)%
\footnote{Formally, for every sequence ${\ov i}=(i_1,...,i_s)\in[N]^s$,
where $N=2^{2cn}$, let $\zeta_{\ov i}(S)$ denote the
output of $D$ when fed with $s_{i_1},...,s_{i_s}$,
where $s_j$ denotes the $j^\xth$ element of the $N$-set $S$.
Then,
$$\mu\eqdef\Exp_S[\zeta_{\ov i}(S)]
 = \prob_{x^{(1)},\dots,x^{(s)}\in\bitset^n}[D(x^{(1)},\dots,x^{(s)})\!=\!1],$$
whereas almost all pairs of $\zeta_{\ov i}(S)$'s are pairwise independent,
because $N=\omega(s^2)$.
Hence,
$$\prob_S\left[\left|\sum_{{\ov i}\in[N]^s}\zeta_{\ov i}(S)-N^s\cdot\mu\right|
               > \beta\cdot N^s \right] \;=\;\frac{O(s^2)}{\beta^2\cdot N}$$
where $s^2/N$ accounts for the fraction of non-disjoint
pairs of ${\ov i}$'s.\label{tedious:fn}}
On the other hand, for every such $2^{2c\cdot n}$-subset $S$,
it holds that $X$ (which is uniform over $S$)
is $\Omega(1)$-far from the uniform distribution over $\bitset^n$
(according to Definition~\ref{dist:def}).
This is the case because $\prob[X\!=\!x]=2^{-2cn}$ for each $x\in S$,
whereas this probability mass has to be transferred
to $2^n/2^{2c n}$ different $n$-bit long strings, but most of these
strings are at relative Hamming distance at least $\e=\Omega(1)$ from $S$
(provided that $\e$ is chosen such that $H_2(\e)<1-2c$).
\EPF

\BOB{\em(Lower bound on testing uniformity over an unspecified support size):}
\label{uniform-unspecified:obs}
For every $c\in(0,0.5)$ there exists $\e>0$ such that 
testing with proximity parameter $\e$ whether a distribution is uniform
over some set requires $2^{c\cdot n}$ samples in the \DHOm.
\EOB

\BPF
We consider the following two families of distributions,
where each of the distributions is parameterized by
an $2^{2c\cdot n}$-subset of $n$-bit strings, denoted $S$.
\begin{enumerate}
\item $X_S$ is uniform on $S$.
\item With probability half, $Y_S$ is uniform on $S$,
and otherwise it is uniform on ${\ov S}\eqdef\bitset^n\setminus S$.
\end{enumerate}
Now, on the one hand, for a random $S$,
no algorithm can distinguish $X_S$ from $Y_S$
by using $o(2^{cn})$ samples (cf.~Footnote~\ref{tedious:fn}).
On the other hand,
we prove that $Y_S$ is far from being uniform on any set.
Suppose that $Y=Y_S$ is $\delta$-close to a distribution
that is uniform on some (arbitrary) set $S'\subseteq\bitset^n$.
We shall show that $\delta=\Omega(1)$,
by considering two cases regarding $S'$:
\BDes
\item[{\em Case 1}:] $|S'|\leq2^{(0.5+c)\cdot n}$ (recall that $c<0.5$).
In this case, the probability mass assigned by $Y$ to ${\ov S}\setminus S'$
should be moved to $S'$, whereas the average relative Hamming distance
between a random element of ${\ov S}\setminus S'$
and the set $S'$ is $\Omega(1)$.
Specifically, letting $U_n$ denote
the uniform distribution on $\bitset^n$,
we upper-bound the probability that $U_n\in{\ov S}\setminus S'$
is H-close to $S'$ by noting that $|{\ov S}\setminus S'|>2^{n-1}$,
since $|S|+|S'|=o(2^n)$,
whereas $|S'|\leq2^{(0.5+c)\cdot n}=2^{n-\Omega(n)}$.
\item[{\em Case 2}:] $|S'|>2^{(0.5+c)\cdot n}$.
In this case, almost all the probability assigned by $Y$ to $S$
should be distributed among more than $2^{(0.5+c)\cdot n}$
strings such that each of these strings is assigned equal weight.
This implies that almost all the weight assigned by $Y$ to $S$
must be moved to strings that are at Hamming distance $\Omega(n)$ from $S$,
since $|S|=2^{2cn}=2^{(0.5+c)\cdot n-\Omega(n)}<2^{-\Omega(n)}\cdot|S'|$.
\EDes
Hence, in both cases, a significant probability weight of $Y$
must be moved to strings that are $\Omega(1)$-H-far from their origin.
The claim follows.
\EPF

\BOB{\em(Lower bound on testing parameterized uniformity, grained,
and support size):}
For every $m\leq 2^{n-\Omega(n)}$,
testing the following properties {\em(equiv., sets)}
of distributions over $\bitset^n$
require $\Omega({\sqrt m})$ samples in the \DHOm:
\BI
\item The set of distributions that are uniform over some $m$-subset;
\item The set of $m$-grained distributions;
\item The set of distributions with support size at most $m$.
\EI
\EOB
Stronger results are presented in
Propositions~\ref{lb:m-support:obs} and~\ref{new-lb:m-grained:prop}.
\medskip

\BPF
As in the proof of Observation~\ref{uniform-on-all:obs},
observe that no algorithm can distinguish
the uniform distribution over $\bitset^n$ from
a distribution that is uniform over an $m$-subset
unless its sees $\Omega({\sqrt m})$ samples.
However, the uniform distribution
over $\bitset^n$ is far from any of the foregoing properties
(also under Definition~\ref{dist:def}), since $m\leq 2^{n-\Omega(n)}$.
\EPF

\BP{\em(lower bound on testing parameterized support size):}
\label{lb:m-support:obs}
For every $m\leq 2^{n-\Omega(n)}$, testing
that a distribution over $\bitset^n$ has support size at most $m$
requires $\Omega(m/\log m)$ samples in the \DHOm.
\EP

\BPF
We use the $\Omega(m/\log m)$ lower bound of~\cite{VV} that
refers to the sample complexity of testing distributions over $[O(m)]$
for support size at most $m$, in the standard testing model
(that is, under the total variation distance).
This lower bound is proved in~\cite{VV} by presenting
two distributions, $X$ and $Y$, that cannot be distinguished
by a {\em label-invariant algorithm}\/ that gets $s=o(m/\log m)$ samples,
where $X$ has support size at most $m$
and $Y$ is far {\em in total variation distance}\/ from
having support size at most $m$.
Using an error correcting code $C:[O(m)]\to\bitset^n$
of constant relative distance, denoted $\delta$,
we consider the distributions $X'=C(X)$ and $Y'=C(Y)$.

On the one hand,
a label-invariant algorithm that obtains $s=o(m/\log m)$ samples
cannot distinguish $X'$ and $Y'$.
Actually, as in the previous proofs, we need to consider
any algorithm that takes $s$ samples, and we identify
for each such algorithm two such distributions $X$ and $Y$
(which are relabelings of the original $X$ and $Y$)
that are indistinguishable by it (cf.~Footnote~\ref{tedious:fn}).
On the other hand, $X'$ has support size at most $m$,
whereas we shall prove that that $Y'$ is far from having
support size at most $m$, under Definition~\ref{dist:def}.
Specifically, we shall show that the hypothesis that $Y$ (and so also $Y'$)
is $\e$-far in total variation distance from having support size at most $m$
implies that $Y'$ is $\Omega(\delta\cdot\e)$-far
(under Definition~\ref{dist:def}) from having support size at most $m$.

Intuitively, $Y'$ is $\Omega(\delta\cdot\e)$-far from having
support size at most $m$, because reducing the support size of $Y'$
requires moving a probability weight of at least $\e$ from
elements in the support of $Y'$ to fewer strings,
whereas the support of $Y'$ consists of strings that
are pairwise $\delta$-far apart in Hamming distance.
Hence, a significant weight of the support of $Y'$
is mapped to a smaller set, which implies that a constant fraction
of this weight must be relocated over a relative Hamming distance
of at least $\delta/2$. Details follow.

Let $Z$ be a distribution that is closest to $Y'$,
under Definition~\ref{dist:def},
among all distributions that are supported on at most $m$ strings,
and let $\gamma$ denote the distance between $Y'$ and $Z$.
By  Definition~\ref{dist:def},
for every ``weight relocation'' function $W :\bitset^{2n} \to [0,1]$
that describes the transformation of $Y'$ to $Z$
(i.e., $W$ satisfies $\sum_{z} W(y',z) = \prob[Y'\!=\!y']$ for every $y' $,
and $\sum_{y'} W(y',z) = \prob[Z\!=\!z]$ for every $z $),
it holds that $\sum_{y'}\sum_{ z} W(y',z) \cdot \Delta_H(y',z) \geq \gamma$;
the former sum is referred to as the \emph{cost} associated with $W$
and it is lower-bounded by the distance between $Y'$ and $Z$
(under Definition~\ref{dist:def}).
Furthermore, $\sum_{y'}\sum_{ z} W(y',z) \cdot \InEq(y',z)$
is lower-bounded by the total variation distance between $Y'$ and $Z$,
where $\InEq(y',z)=1$ if $y'\neq z$ and $\InEq(y',y')=0$.
We shall show that $\gamma$ is lower-bounded
by the total variation (TV) distance between $Y'$ and $Z$,
which in turn is lower-bounded by TV-distance of $Y'$ from
having support size at least $m$.

Let $S$ denote the support of $Z$ (so that $W(y',z)=0$ for every $z\notin S$),
and let $S'$ be the subset of $S$ that contains 
strings that are $(0.4\cdot\delta)$-H-close to the code $C$.
Recall that the support of $Y'$ is a subset of $C$
(so that $W(y',z)=0$ for every $y'\notin C$).
Using $S=(S\setminus S')\cup (S'\setminus C)\cup (S'\cap C)$,
the cost associated with $W$ is the sum of following three sums.
\begin{eqnarray}
\label{cost-of-W:sum1:eq}
& & \sum_{y'}\sum_{ z\in S\setminus S'} W(y',z)\cdot \Delta_H(y',z) \\
\label{cost-of-W:sum2:eq}
& & \sum_{y'}\sum_{ z\in S'\setminus C} W(y',z)\cdot \Delta_H(y',z) \\
& & \sum_{y'}\sum_{ z\in S'\cap C} W(y',z)\cdot \Delta_H(y',z).
\label{cost-of-W:sum3:eq}
\end{eqnarray}
We analyze each sum separately,
while letting $R$ denote the support of $Y'$.
\BI
\item
By the definition of $S'$
(and since the support of $Y'$ is a subset of $C$),
for each $y'$ in the support of $Y'$ and each $z\in S\setminus S'$,
we have that $\Delta_H(y',z) > 0.4\cdot\delta$.
Therefore, \eqref{cost-of-W:sum1:eq} is lower-bounded by
$\sum_{y'}\sum_{ z\in S\setminus S'} W(y',z)\cdot 0.4\cdot\delta$.

\item Turning \eqref{cost-of-W:sum2:eq}, for each $z\in S'\setminus C$,
let $\ccode(z) \in C$ be the codeword in $C$ that is closest to $z$.
By the definition of $S'$ we have that
$\delta'(z) \eqdef \Delta_H(\ccode(z),z) \leq 0.4\cdot \delta$,
and for every $y'\in R\setminus\{\ccode(z)\}$,
we have that $\Delta_H(y',z) \geq \delta - \delta'(z) \geq 0.6\cdot \delta$.

We claim that (for every $z\in S'\setminus C$),
at least half the probability mass that is relocated by~$W$ (from $Y'$)
to $z$ must come from codewords $y'$ (in the support of $Y'$)
that are different from $\ccode(z)$;
that is,
\begin{equation}
\label{sum2-charging-rule:eq}
\sum_{y'\in R\setminus\{\ccode(z)\}} W(y',z)
  \geq \frac{1}{2}\cdot \sum_{y'} W(y',z).
\end{equation}
We shall actualy prove that
$\sum_{y'\in R\setminus\{\ccode(z)\}} W(y',z) \geq W(\ccode(z),z)$,
and do so by showing that otherwise we could modify $Z$ (and $W$)
to obtain a distribution $Z'$ with support size at most $m$
(and a corresponding weight relocation function $W'$)
such that $Z'$ is closer to $Y'$ than $Z$.
Indeed,
a contradiction follows when assuming (w.l.o.g.) that $W$ is optimal
(i.e., it has the lowest cost among all relocation functions
that transform $Y'$ into $Z$, which in turn was defined
as a closest to $Y'$ distribution of support size at most $m$).

Specifically, suppose towards the contradiction
that some $z\in S'\setminus C$
satisfies $W(\ccode(z),z) > \sum_{y'\in R\setminus\{\ccode(z)\}} W(y',z)$.
Then, consider $Z'$ obtained by moving the probability mass
that~$Z$ assigns $z$ to the codeword $\ccode(z)$;
that is, $\prob[Z'\!=\!z] = 0$
and $\prob[Z'\!=\!\ccode(z)]=\prob[Z\!=\!\ccode(z)]+\prob[Z\!=\!z]$
(and $\prob[Z'\!=\!z']=\prob[Z\!=\!z']$ for every $z'\notin\{z,\ccode(z)\}$),
while noting that~$Z'$ has support size at most $m$.
The weight relocation function $W'$ is define accordingly
(i.e., for each $y'$, we set $W'(y',z)=0$
and $W'(y',\ccode(z)) = W(y',\ccode(z))+W(y',z)$
(leaving $W'(y',z') = W(y',z')$ for every $z' \notin \{z,\ccode(z)\}$)).
We observe that the cost of $W'$
(which upper-bounds the distance between $Y'$ and $Z'$)
equals the cost of $W$ minus $\sum_{y',z}W(y')\cdot \Delta_H(y',z)$
plus $\sum_{y'}W(y',z)\cdot \Delta_H(y',\ccode(z))$.
Now, observe that
\begin{equation}
\label{lb:m-support:gain:eq}
\sum_{y'\in R}W(y',z)\cdot \Delta_H(y',z)
   \geq W(\ccode(z),z)\cdot \delta'(z)
   +\sum_{y'\in R\setminus\{\ccode(z)\}} W(y',z)\cdot (\delta - \delta'(z))\,,
\end{equation}
since for $y'\in R\setminus\{\ccode(z)\}$ it holds that
$\Delta_H(y',z) \geq \Delta_H(y',\ccode(z))-\Delta_H(z,\ccode(z))
 \geq \delta-\delta'(z)$,
whereas
\begin{equation}
\label{lb:m-support:loss:eq}
\sum_{y'\in R}W(y',z)\cdot \Delta_H(y',\ccode(z))
   \leq  \sum_{y'\in R\setminus\{\ccode(z)\}} W(y',z)\cdot \delta\,.
\end{equation}
Using the contradiction hypothesis
(i.e., $W(\ccode(z),z) > \sum_{y'\in R\setminus\{\ccode(z)\}} W(y',z)$),
we infer that the l.h.s of \eqref{lb:m-support:gain:eq}
is larger than the l.h.s of \eqref{lb:m-support:loss:eq},
and reach a contradiction to the optimality of $W$
(since the cost of $W'$ is smaller than the cost of $W$).

Hence, we have established that, for each $z\in S'\setminus C$,
\eqref{sum2-charging-rule:eq} holds; that is,
$$\sum_{y'\in R\setminus\{\ccode(z)\}} W(y',z)
  \geq \frac{1}{2}\cdot \sum_{y'} W(y',z).$$
Recalling that $\Delta_H(y',z) \geq 0.6\cdot \delta$,
it follows that
$$\sum_{y'\in R} W(y',z)\cdot\Delta_H(y',z)
  \geq \frac{1}{2}\cdot \sum_{y'} W(y',z) \cdot 0.6\cdot \delta,$$
which implies that \eqref{cost-of-W:sum2:eq} is lower-bounded by
$\sum_{y'}\sum_{z\in S'\setminus C} W(y',z)\cdot 0.3\cdot\delta$.

\item Lastly we consider \eqref{cost-of-W:sum3:eq}.
Clearly, for each $y'$ in the support of $Y'$
and each $z\in S'\cap C$ such that $z\neq y'$,
we have that $\Delta_H(y',z) \geq \delta$.
Therefore, \eqref{cost-of-W:sum3:eq} is lower-bounded
by $\sum_{y'}\sum_{z\in (S'\cap C)\setminus \{y'\}} W(y',z)\cdot \delta$,
which we rewrite
as $\sum_{y'}\sum_{z\in(S'\cap C)} W(y',z)\cdot\InEq(y',z)\cdot\delta$.
\EI
Note that $\InEq(y',z)=1$
for every $y'\in R$ and $z\in(S\setminus S')\cup(S'\cup C)$.
Combining the lower bounds
on \eqref{cost-of-W:sum1:eq}--(\ref{cost-of-W:sum3:eq}),
it follows that the cost of $W$ is
at least $\sum_{y',z}W(y',z)\cdot\InEq(y',z)\cdot0.3\delta$.
Hence, the distance (denoted $\gamma$) between $Y'$ and $Z$,
under Definition~\ref{dist:def},
is at least a $0.3\delta$ factor of the total variation distance
between these two distributions.
\EPF

\BP{\em(Lower bound on testing $m$-grained distributions):}
\label{new-lb:m-grained:prop}
For every constant $c<1$ and $m\leq 2^{n-\Omega(n)}$,
testing that a distribution over $\bitset^n$ is $m$-grained
requires $\Omega(m^c)$ samples in the \DHOm.
\EP
We comment that the foregoing lower bound (for \DHOm)
matches the best known lower bound
for the standard distribution-testing model~\cite{GR:gr}.
See Section~\ref{conditional-lb:sec} for further discussion.
\medskip

\BPF
We use the $\Omega(m^c)$ lower bound of~\cite{GR:gr}
that refers to he sample complexity of testing
whether a distribution over $[O(m)]$ is $m$-grained,
in the standard model (i.e., under the total variation distance).
This lower bound is proved in~\cite{GR:gr} by presenting
two ($2m$-grained) distributions, $X$ and $Y$,
that cannot be distinguished by a label-invariant algorithm
that gets $s=o(m^{c})$ samples, where~$X$ is $m$-grained
and $Y$ is far (in total variation distance) from being $m$-grained.

As in the proof of Proposition~\ref{lb:m-support:obs},
applying an error correcting code $C:[O(m)]\to\bitset^n$
to $X$ and $Y$, we observe that $X'=C(X)$ is $m$-grained
whereas $Y'=C(Y)$ is far from being $m$-grained
(also under Definition~\ref{dist:def}).%
\footnote{In fact, as in the proof of Proposition~\ref{lb:m-support:obs},
we actually consider adequate relabelings of $X$ and $Y$.}
To see that $Y'$ is far from any distribution $Z$
that is $m$-grained and is supported by a set $S$,
we (define $S'$ and) employ the same case-analysis
as in the proof of Proposition~\ref{lb:m-support:obs}.
Note that in the second case
(i.e., probability mass relocated from $Y'$ to $z\in S'\setminus C$),
the potential replacement (of $z$ by the codeword closest to it)
preserves $m$-grained-ness.
Hence, as in the proof of Proposition~\ref{lb:m-support:obs},
the distance (under Definition~\ref{dist:def}) between $Y'$ and $Z$
is lower-bounded by a constant fraction of their total variation distance.
\EPF

\BP{\em(Lower bound on testing parameterized uniformity):}
\label{new-lb:m-uniformity:prop}
For every constant $c<1$ and $m\leq n$,
testing that a distribution over $\bitset^n$ is uniform over some $m$-subset
requires $\Omega(m^c)$ queries in the \DHOm.
\EP
We stress that, unlike Proposition~\ref{new-lb:m-grained:prop},
which lower-bounds the sample complexity of testers,
in Proposition~\ref{new-lb:m-uniformity:prop} we only lower-bound
their query complexity.%
\footnote{We actually use $m\log m = o(n^{1/c})$,
which follows from $m\leq n$.
This inequality is used in order to derive the simple form
of the lower bound from
the actual lower bound of $\Omega(\min(m^c,n/\log m))$.}
\medskip

\BPF
Let $X'$ and $Y'$ denote the distributions
derived in the proof of Proposition~\ref{new-lb:m-grained:prop}.
Recall that $X'$ is $m$-grained,
whereas $Y'$ is $\Omega(1)$-far from being $m$-grained
(under Definition~\ref{dist:def}).
Note that $Y'$ is $\Omega(1)$-far from being uniform over
any set of size $m$,
and observe that $X'$ is $\frac{\log_2m}{n}$-close
to a distribution $X''$ that is uniform over a set of size $m$.
Specifically, we can transform $X'$ to $X''$
by modifying only the bits that reside in $\log_2m$ locations,
where the choice of these locations is arbitrary.%
\footnote{Saying that $X'$ is $m$-grained means that it is uniform
on a multi-set $\{x^{(1)},\dots,x^{(m)}\}$ of $n$-bit strings.
We modify $X'$ by replacing each $x^{(i)}$ by $y^{(i)}$
such that $y^{(i)}$ encodes the binary expansion of $i-1$
in the chosen locations and equals~$x^{(i)}$ otherwise.
That is, letting $\ell_1<\ell_2<\cdots<\ell_{\log_2m}$
denote the chosen locations, we set $y^{(i)}_{\ell_j}$
to equal the $j^\xth$ bit in the binary expansion of $i-1$
and set $y^{(i)}_\ell=x^{(i)}_\ell$ if
$\ell\in[n]\setminus\{\ell_1,\ell_2,\dots,\ell_{\log_2m}\}$.}
Hence, a potential tester that make $o(n/\log m)$ queries
is unlikely to hit these locations,
if we select these locations uniformly at random.
Using $m\leq n$, we conclude that a potential tester
that makes $\min(o(m^c),o(n/\log m))=o(m^c)$ queries
cannot distinguish between the distribution $X''$
and distribution $Y'$, which implies that it fails
to test uniformity in the \DHO\ model.
\EPF

\subsection{Conditional lower bounds}
\label{conditional-lb:sec}
The lower bounds (for the \DHOm) presented in
Propositions~\ref{new-lb:m-grained:prop} and~\ref{new-lb:m-uniformity:prop}
build on the best known lower bound for testing the set of grained
distributions in the standard distribution-testing model.
Improved lower bounds on the complexity of testing in the \DHOm\
(see Theorems~\ref{lb:m-grained:thm} and~\ref{lb:m-uniformity:thm})
rely on a conjecture regarding the sample complexity
of testing grained distributions in the standard model.

\BCJ{\em(On the complexity of testing the set of $m$-grained distributions
in the standard distribution-testing model):}
\label{grained:conj}
In the standard distribution-testing model,
the sample complexity of testing $m$-grained distributions
over a domain of size $O(m)$ is $\Omega(m/\log m)$.
\ECJ

\BT{\em(On testing the set of $m$-grained distributions in the \DHOm):}
\label{lb:m-grained:thm}
Assuming Conjecture~\ref{grained:conj},
for every $m\leq 2^{n-\Omega(n)}$,
testing  that a distribution over $\bitset^n$ is $m$-grained
requires $\Omega(m/\log m)$ samples in the \DHOm.
\ET
We mention that the proof would remain unchanged if
the lower bound in Conjecture~\ref{grained:conj}
is replaced by $s(m)$; that is,
any lower bound of the form $\Omega(s(m))$ on the sample complexity
of testing $m$-grained distributions in the standard distribution
testing model translates to an $\Omega(s(m))$ lower bound in the \DHOm.
A similar comment refers to Theorem~\ref{lb:m-uniformity:thm}.
\medskip

\BPF
We would have liked to argue that the proof is analogous to
the proof of Proposition~\ref{new-lb:m-grained:prop},
except that here we assume the existence of two distributions, $X$ and $Y$,
over $[O(m)]$ that cannot be distinguished by a label-invariant
algorithm that gets $o(m/\log m)$ samples, where~$X$ is $m$-grained
and $Y$ is far (in total variation distance) from being $m$-grained.
However, since Conjecture~\ref{grained:conj} does not quite
imply the existence of such distributions $X$ and $Y$,
we apply a slightly more complex argument.
Our starting point is the observation that Conjecture~\ref{grained:conj}
implies the existence of multi-sets of distributions%
\footnote{Actually, $\X$ and $\Y$ are distributions of distributions.
However, to avoid confusion, we preferred to present them
as multi-set and consider a uniformly selected element in them.},
denoted $\X$ and $\Y$, such that the following holds:
\begin{enumerate} 
\item Each distribution in $\X$ is $m$-grained.
\item Each distribution in $\Y$ is TV-far from being $m$-grained.
\item No algorithm can distinguish between $s=o(m/\log m)$ samples
taken from a distribution~$X$ that is selected uniformly in $\X$
and $s$ samples taken from a distribution $Y$
that is selected uniformly in $\Y$.
\end{enumerate} 
The foregoing observation is proved by applying the MiniMax Principle
(cf.~\cite[Apdx~A.1]{G:mi-pt}).
Specifically, we consider deterministic algorithms that,
given $s$ samples from a distribution~$Z$, try to distinguish
between the case that $Z$ is $m$-grained and the case that $Z$
is TV-far from being $m$-grained, and denote by $c(A,Z)$
the probability that algorithm $A$ is correct on $Z$
(i.e., it correctly identifies $Z$'s type).
Then, Conjecture~\ref{grained:conj} asserts that,
for every distribution $\A$ of algorithms
(i.e., a randomized algorithm) that get $s$ samples,
there exists a distribution $Z$
(which is either $m$-grained or far from $m$-grained)
such that $\A$ errs on $Z$ with probability greater than~$1/3$
(i.e., $\Exp_{A\sim\A}[c(A,Z)]<2/3$).
Applying the minimax principle, it follows that there exists
a multi-set $\ZZ$ of such distributions
(which are each either $m$-grained or far from $m$-grained)
on which each algorithm $A$ that takes $s$ samples errs
on the average with probability greater than~$1/3$
(i.e., $\Exp_{Z\in\ZZ}[c(A,Z)]<2/3$).
Analogously to~\cite[Exer.~7.3]{G:pt},
we obtain $\X$ and $\Y$ as desired,
where the indistinguishability gap is less than~$1/2$.

Consider the corresponding multi-sets $\X'$ and $\Y'$,
which are obtained by applying a (constant-distance) error correcting code $C$
to the elements of each distribution in $\X$ and $\Y$, respectively.
Then, no algorithm that takes $s$ samples
can distinguish $X'$ from $Y'$, where $X'$ (resp.,~$Y'$)
is selected uniformly in $\X'$ (resp., $\Y'$),
where the indistinguishability gap is less than~$1/2$.
(As in the proof of Proposition~\ref{new-lb:m-grained:prop},
one can show that each $Y'$ is far
(under Definition~\ref{dist:def}) from being $m$-grained.)
Observing that a distinguishing gap of less than~$1/2$ means that
no algorithm (of low complexity) constitutes a tester with error
probability at most~$1/4$ (rather than at most~$1/3$),
the claim follows (by using error reduction).
\EPF

\BT{\em(On testing parameterized uniformity in the \DHO\ model):}
\label{lb:m-uniformity:thm}
Assuming Conjecture~\ref{grained:conj},
for every $m\leq n$, testing
that a distribution over $\bitset^n$ is uniform over some $m$-subset
requires $\Omega(m/\log m)$ queries in the \DHOm.
\ET
We stress that, unlike Theorem~\ref{lb:m-grained:thm},
which lower-bounds the sample complexity of testers,
in Theorem~\ref{lb:m-uniformity:thm} we only lower-bound
their query complexity.
\medskip

\BPF
Let $\X'$ and $\Y'$ denote the multi-sets of distributions
derived in the proof of Theorem~\ref{lb:m-grained:thm}.
Recall that each distribution in $\X'$ is $m$-grained,
whereas each distribution in $\Y'$ is far from being $m$-grained
(under Definition~\ref{dist:def}).
As in the proof of Proposition~\ref{new-lb:m-uniformity:prop},
note that each distribution in $\Y'$ is $\Omega(1)$-far from
being uniform over a set of size $m$,
and observe that each distribution in $\X'$ is $\frac{\log_2m}{n}$-close
to being uniform over a set of size $m$.
Specifically, we can make each distribution in $\X'$ uniform
by modifying only the bits that reside in $\log_2m$ locations,
where the choice of these locations is arbitrary.
Hence, a potential tester that make $o(n/\log m)$ queries
is unlikely to hit these locations,
if we select these locations uniformly at random.
Using $m\leq n$,
we conclude that a potential tester that makes $o(m/\log m)$ queries
cannot distinguish between distributions in the modified multi-set $\X'$
and distributions in the multi-set $\Y'$, which implies that it fails
to test uniformity in the \DHO\ model.
\EPF

\section{Distributions on Self-Correctable/Testable Sets}
\label{selt-correction:sec}
In this section we prove Theorem~\ref{tentative:ithm},
which refers to properties of distributions
that are supported on a set of strings $\Pi\subseteq\bitset^n$
that has an efficient self-correction/testing procedure.
In addition, we postulate that these distributions are
label-invariant when restricted to $\Pi$; that is,
for every bijection $\pi:\Pi\to\Pi$ and every distribution $X$,
it holds that $X$ is in the property
if and only if $\pi(X)$ is in the property.

Our starting point is a label-invariant property of distributions,
denoted $\D$, and a property of strings, denoted $\Pi$,
that has a relatively efficient tester and local self-corrector.
Actually, we use a {\em relaxed definition of self-correction},
which allows to output a special failure symbol in
case the input (oracle) is not in $\Pi$ (but is close to $\Pi$).%
\footnote{The notion of relaxed self-correction was introduced in~\cite{GRR},
by analogy to the notion of relaxed LDCs~\cite[Sec.~4.2]{BGHSV}.}
Indeed, proper behavior of the self-corrector is required only
up to a specified distance from the set $\Pi$.
Combining $\D$ and $\Pi$, we get a property of distributions,
denoted $\D_\Pi$, that consists of all distributions in $\D$
that are supported by $\Pi$.
Note that $\D_\Pi$ is {\em label-invariant over its support},
because for every $X$ in $\D_\Pi$ and every
bijection $\pi:\Pi\to\Pi$ it holds that $\pi(X)$
is in $\D$ and has support $\Pi$.
We prove that the query complexity of testing $\D_\Pi$
in the \DHO\ model 
is related to the sample complexity of testing $\D$
(in the standard model) and to the query complexity
of the two aforementioned procedures.

\BT{\em(From standard distribution testing of $\D$ to testing $\D_\Pi$ in
the \DHO\ model, when~$\Pi$ is efficiently testable and self-correctable):}
\label{li+sc:thm}
\BI
\item
Let $\D$ be a label-invariant property of distributions over $\bitset^n$,
and suppose that $\D$ is 
testable in the standard 
model
using $s(n,\e)=\Omega(1/\e)$ samples.%
\footnote{Indeed, it would have been more consistent
with the literature to denote the sample complexity
by $s(2^n,\e)$, since the domain size is $2^n$.}
\item
Let $\Pi\subseteq\bitset^n$ be a property of strings
that is testable with query complexity $\qt(n,\e)$
and self-correctable up to distance $\delta(n)$
with $\qc(n)$ queries; that is, there exists
an oracle machine~$C$ that makes at most $\qc(n)$ queries
such that for every $x\in\bitset^n$ and $i\in[n]$
the following two conditions hold:
\begin{enumerate}
\item If $x\in\Pi$, then $\prob[C^{x}(i)\!=\!x_i]\geq2/3$.
\item If $x$ is $\delta(n)$-close to $x'\in\Pi$,
then $\prob[C^{x}(i)\!\in\!\{x'_i,\bot\}]\geq2/3$,
where $\bot$ is a special symbol indicating failure.
\end{enumerate}
\item
Suppose that every distribution in $\D$ is supported
by a subset of size at most $|\Pi|$,
and  let $\DP$ denote the set of all distributions in $\D$ that have
a support that is a subset of $\Pi$.%
\footnote{The condition
(regarding the size of the support of distributions in $\D$)
implies that {\em if $X'$ is supported
by a subset of $\Pi$ and is TV-close to $\D$,
then it is TV-close to $\DP$}.
Note that this implication does not hold without the condition.
(Consider, for example, $\Pi=\{0^n,1^n\}$ and $\D$ that contains
all distributions that are uniform on some 2-subset as
well as all distribution that have support size $2^n$.
Then, $X'\equiv0^n$ is arbitrarily TV-close to $\D$,
although $X'$ is TV-far from $\DP$
(which consists of the distribution that is uniform on $\{0^n,1^n\}$).}
\item
Suppose that $\qt(n,\e)\geq2\cdot\qt(n,2\e)$ holds
for all $\e\in[O(1/n),\Omega(1)]$.%
\footnote{This natural assumption is made in order to save
a factor of $1/\e$ in \eqref{li+sc:complexity:eq}.}
\EI
Then, $\DP$ is testable with query complexity
\begin{equation}
\label{li+sc:complexity:eq}
q(n,\e)=\tildeO(s(n,\e'/2))
  \cdot\frac{\qt(n,\delta(n))+\qc(n)}{\delta(n)}
\end{equation}
and sample complexity $s(n,\e'/2)$, where $\e'=\min(\e,\delta(n))$.
\ET
Note that $\Pi$ must have relative (Hamming) distance greater $\delta(n)$,
since otherwise we reach a contradiction by considering
two strings in $\Pi$ that are $\delta(n)$-H-close to one another.%
\footnote{Let $x$ and $x'$ be two distinct strings that are
at relative distance at most $\delta(n)$, and suppose than $x_i\neq x'_i$.
Then, by the first condition $\prob[C^x(i)\!=\!x_i]\geq2/3$, but by
the second condition $\prob[C^x(i)\!\in\!\{x'_i,\bot\}]\geq2/3$
(since $x$ is $\delta(n)$-close to $x'$).}

In some natural applications
(e.g., properties of distributions over low-degree multi-variate polynomials),
we may use $\delta(n)\geq1/\poly(\log n)$
and $\qt(n,\delta(n))+\qc(n)\leq\poly(\log n)$.
When testing properties of distributions over a
relaxed locally correctable (and testable) code~\cite{BGHSV,GRR,CGS},
we may use $\delta(n)=\Omega(1)$ and $\qt(n,\delta(n))+\qc(n)=O(1)$.
\medskip

\BPF
We first recall that, we may assume, without loss of generality,
that when testing a label-invariant property, in the standard model,
the tester is label-invariant~\cite{Batu}
(see also~\cite[Thm.~11.12]{G:pt}). 
Such a tester, denoted $T'$, actually rules according to
the collision pattern that it sees in the sequence of samples; that is,
the number of $i$-way collisions in the sequence, for each $i\in\N$.
Specifically, the {\sf collision pattern}
of the sequence ${\ov x}=(x^{(1)},\dots,x^{(s)})$, denoted $\CP({\ov x})$,
is the sequence $(c_1,\dots,c_s)$ such that
$c_i=|\{v\!\in\!\Omega:\#_v(x^{(1)},\dots,x^{(s)})=i\}|$
is the number of $i$-way collisions,
where $\#_v(x^{(1)},\dots,x^{(s)})=|\{j\in[s]:x^{(j)}=v\}|$
is the number of elements (in the sequence) that equal $v$.
Hence, we claim that
$$T'(n,\e;x^{(1)},\dots,x^{(s)})=T''(n,\e;\CP(x^{(1)},\dots,x^{(s)})),$$
for some randomized decision procedure $T''$.

\myparagraph{Warm-up (or a first attempt).}
Using the fact that $\Pi$ has relative distance $\delta=\delta(n)$,
let us consider a tester that, given
the sample-sequence ${\ov x}=(x^{(1)},\dots,x^{(s)})$, where $s=s(n,\e)$,
picks uniformly at random an $O(\delta^{-1}\log s)$-subset of $[n]$,
denoted $I$, and outputs $T''(n,\e;\CP(x^{(1)}_I,\dots,x^{(s)}_I))$,
where $x^{(i)}_I$ is the 
restriction of $x^{(i)}$ 
to the coordinates in $I$.
This tester works well when the tested distribution $X$
is supported on $\Pi$:
In this case,
for any sequence of $s$ samples ${\ov x}=(x^{(1)},\dots,x^{(s)})$ of $X$,
with high probability, $\CP(x^{(1)}_I,\dots,x^{(s)}_I)$
equals $\CP(x^{(1)},\dots,x^{(s)})$, since $x^{(i)}\neq x^{(j)}$ implies
$\prob_I[x^{(i)}_I\!=\!x^{(j)}_I]<(1-\delta)^{O(\delta^{-1}\log s)}=o(1/s^2)$.
Hence, in this case we correctly distinguish $X$ in $\DP$ from $X$
that is $\e$-far from $\DP$ although it is supported by $\Pi$.
Of course, we can easily test that $X$ is supported by $\Pi$
(using the tester for $\Pi$ -- see Theorem~\ref{support:ithm}), but the problem is that our samples
may be close to $\Pi$ and yet not reside in it.
This is a problem because the foregoing analysis presupposed
that the inequality between samples is reflected in their
restrictions to a small subset $I$ (i.e., that $x^{(i)}\neq x^{(j)}$
typically implies $x^{(i)}_I\!=\!x^{(j)}_I]$).

We address 
this problem by using the 
hypothesis regarding $\Pi$; that is, not only is $\Pi$  
testable (with proximity parameter $\e$ using $\qt(n,\e)$  queries), but it is also self-correctable
to distance~$\delta$ by using  $\qc(n)$ queries. 
In particular, combining the 
tester for $\Pi$ (applied with proximity parameter~$\delta(n)$),
and the self-corrector (and employing error reduction)%
\footnote{In the context of self-correction,
performing error reduction means that if a strict majority
of the invocations return a Boolean value,
then we can use that value (which happens w.h.p.\/ when $x\in\Pi$).
Otherwise, we output $\bot$,
since the lack of a strict majority indicates error.}
%
we can obtain an oracle machine $C$ that satisfies
the following for every $x\in\bitset^n$ and $i\in[n]$
(for any integer parameter $s$):
\begin{enumerate}
\item If $x\in\Pi$, then $\prob[C^x(i)\!=\!x_i]\geq1-o(1/s^2)$.
\item If $x$ is $\delta$-H-close to $x'\in\Pi$,
then $\prob[C^x(i)\!\in\!\{x'_i,\bot\}]\geq1-o(1/s^2)$.
\item If $x$ is $\delta$-H-far from $\Pi$,
then $\prob[C^x(i)\!=\!\bot]\geq1-o(1/s^2)$.
\end{enumerate}
This combined machine has query complexity
$\qcp(n,s)=O(\log s)\cdot(\qt(n,\delta)+\qc(n))$.
We are now ready to present an analyze our tester.


\myparagraph{The proposed tester, $T$.}
On input $n,\e$ and $s=s(n,\e/2)$ samples
drawn from a tested distribution $X$,
and denoted $x^{(1)},\dots,x^{(s)}$, we proceeds as follows
(assuming, without loss of generality, that $\e\leq\delta(n)/2$).

\begin{enumerate}
\item
We test whether the distribution $X$ is supported by $\Pi$,
with proximity parameter $\e/2$,
where the distance here
(and throughout the proof, unless explicitly stated otherwise)
is according to Definition~\ref{dist:def}.
If this test rejects, then we reject.

(Here and below, ``supported by $\Pi$'' means
having a support that is a subset of $\Pi$.)

The aforementioned testing is performed by using
the distribution-tester provided by Theorem~\ref{support:thm}
(with proximity parameter $\e/2$),
while noting that we can reuse some of the samples provided to $T$
for this purpose (since $s=\Omega(1/\e)$).
The query complexity of this distribution-tester
is $\poly(\log(1/\e))\cdot\qtp(n,\e)$,
where $\qtp(n,\e)=O(\qt(n,\delta(n)))+\tildeO(1/\e)\cdot\qc(n)$
is the query complexity of testing $\Pi$ as follows:
On input $x$, we first invoke the original tester
with proximity parameter set to $\delta(n)$,
and then select $O(1/\e)$ random locations $i\in[n]$ and compare their
self-corrected value (i.e., $C^x(i)$) to their given value (i.e., $x_i$).%
\footnote{This idea is implicit in the proof of~\cite[Thm.~5.12]{G:pt}.
The extra $O(\log(1/\e))$ factor accounts for reducing the
error of the self-corrector to $o(1/\e)$.}

\item
For each $i\in[s]$, we test whether $x^{(i)}$ is in $\Pi$,
when setting the proximity parameter to $\delta=\delta(n)$
and the error bound to $o(1/s)$.
If any of these checks rejects, then we reject.
Otherwise, we may assume that each sample
is $\delta(n)$-H-close to $\Pi$.

The query complexity of the current step is $\tildeO(s)\cdot\qt(n,\delta(n))$.
Note that typically $s\gg 1/\e$ and $\delta\gg\e$,
which implies that the complexity of the current step
is incomparable to that of Step~1.%
\footnote{On the one hand, we test that all samples are in $\Pi$
(rather than only testing the first $O(1/\e)$ samples).
On the other hand, the proximity parameter we use is larger;
that is, the test is more crude.}
%
\item
For $\ell=O(\delta^{-1}\log s)$,
select a random $\ell$-subset of $[n]$, denoted $I$,
and obtain $(y^{(1)},\dots,y^{(s)})$ such that $y^{(i)}$
is the self-correction of $x^{(i)}$ for locations $I$;
that is, letting $I=\{i_1,\dots,i_\ell\}$
such that $i_j<i_{j+1}$ (for all $j\in[s-1]$),
we let $y^{(i)}_j\gets C^{x^{(i)}}(i_j)$.
(Recall that $C$ has error probability $o(1/s^2)$.)

If any of these correction attempts fails
(i.e., if any $y^{(i)}_j$ equals $\bot$), then we reject.
Otherwise, we output the verdict of $T''(n,\e/2;\CP(y^{(1)},\dots,y^{(s)}))$.
We may assume, without loss of generality,
that $T'$ (and so $T''$) errs with probability at most $0.1$.

The query complexity of the current step is
$s\cdot\ell\cdot\qcp(n,s)
  = O(s\cdot\log^2s)\cdot\frac{\qt(n,\delta(n))+\qc(n)}{\delta(n)}$.
\end{enumerate}
The query complexity of $T$ is upper-bounded by
$$\left(\tildeO(1/\e)+\tildeO(s)+\frac{\tildeO(s)}{\delta(n)}\right)
   \cdot\left(\qt(n,\delta(n))+\qc(n)\right)
  = \tildeO(s)\cdot\frac{\qt(n,\delta(n))+\qc(n)}{\delta(n)}$$
where the last equality uses $s=\Omega(1/\e)$.

\myparagraph{Analysis of the proposed tester $T$.}
We start with the case that $X$ is in $\DP$.
In this case, the first two steps cause rejection
with probability $o(1)$, since all $x^{(i)}$'s are in $\Pi$.
Furthermore, in this case, with probability $1-o(1)$,
each $y^{(i)}$ equals the 
restriction of $x^{(i)}$ 
to the locations in~$I$
(i.e., $y^{(i)}=x^{(i)}_I$).
As argued in the motivational discussion, if $x^{(i)}\neq x^{(j)}$,
then $x^{(i)}$ and~$x^{(j)}$ are $\delta$-H-far apart from one another,
and so $\prob_{p\in[n]}[x^{(i)}_p=x^{(j)}_p]<1-\delta$.
This implies that $\prob_{I}[x^{(i)}_I=x^{(j)}_I]<(1-\delta)^\ell=o(1/s^2)$,
by our choice of $\ell$.
We conclude that
$\prob_I[\CP(y^{(1)},\dots,y^{(s)}))\!=\!\CP(x^{(1)},\dots,x^{(s)}))]=1-o(1)$,
which implies that our tester accepts $X$
with probability at least $0.9-o(1)$.

We now turn to the case that $X$ is $\e$-far from $\DP$
(according to Definition~\ref{dist:def}).
The easy case is that $X$ is $\e/2$-far from being supported by $\Pi$,
and this case leads Step~1 to reject with very high probability.
We thus assume that $X$ is $\e/2$-close to being supported by $\Pi$,
and let $\DEC(x)$ denotes the string in $\Pi$ that is closest to $x$.
Then, in expectation, $\DEC(X)$ is $\e/2$-H-close to $X$,
since the Hamming distance between~$x$ and $\DEC(x)$
equals the Hamming distance between~$x$ and $\Pi$.
Hence, $\DEC(X)$ is $\e/2$-close to $X$,
which implies that $\DEC(X)$ is $\e/2$-far from $\DP$.
By the next claim, this implies that
the total variation distance between $\DEC(X)$
and $\D$ is greater than~$\e/2$.

\Bcm{\em(Distance to $\DP$ vs TV-distance to $\D$):}
\label{li+sc:tvd-clm}
Let $X'$ be a distribution supported by $\Pi$
such that $X'$ is $\e'$-far from $\DP$
{\rm(according to Definition~\ref{dist:def})}.
Then, the total variation distance between~$X'$ 
and $\D$ is greater than $\e'$.
\Ecm

\Bpf
Assume, contrary to the claim,
that $X'$ is $\e'$-TV-close to some distribution $Y'$ in $\D$.
If $Y'$ in~$\DP$, then we immediately reach a contradiction
to the hypothesis of the claim by which $X'$ is
at distance greater than $\e'$ from $\DP$
(according to Definition~\ref{dist:def}).
This is the case because (as noted in the introduction),
the total variation distance between distributions
upper-bounds the distance according to Definition~\ref{dist:def}.
Hence, $Y'$ in $\D\setminus\DP$.
We claim that in such a case, using $Y'$,
we can define a distribution $Y''$ in $\DP$ such that $X'$
is $\e'$-TV-close to $Y''$, resulting once again in a contradiction.
Thus, it remains to establish the existence of such a distribution $Y''$.

Recall that by the premise of Theorem~\ref{li+sc:thm},
the support size of $Y'$ is at most $|\Pi|$.
Let $S'$ denote the support of $Y'$, and $S'' = S'\setminus \Pi$.
Consider any subset $\Pi''$ of $\Pi\setminus S'$ such that $|\Pi''| = |S''|$
(such a subset must exist because $|S'| \leq |\Pi|$ and hence
$|S''|  = |S' \setminus \Pi| \leq |\Pi\setminus  S'|$).
Selecting any bijection $\phi$ between $S''$ and $\Pi''$,
we set $\prob[Y''=j] = \prob[Y' = \phi^{-1}(j)]$ for every $j \in \Pi''$,
and $\prob[Y''=j] = \prob[Y'=j]$ for every $j \in \Pi\cap S'$.
Note that the total variation distance between~$X'$ and~$Y''$
is upper-bounded by the total variation between $X'$ and $Y'$,
because the probability mass assigned by $Y''$ to $\Pi''$
is already charged to the TV-distance between $Y'$ and $X'$
(since $\prob[Y''\!\in\!\Pi'']=\prob[Y'\!\in\!S'']$
whereas $\prob[X'\!\in\!S'']=0$).
\Epf


\medskip
By applying Claim~\ref{li+sc:tvd-clm}, we get that the total variation distance
between $\DEC(X)$ and $\D$ is greater than $\e/2$.
It follows that, with probability at least~0.9,
the 
(standard) tester for $\D$ (i.e.,~$T'$), 
rejects when given $s=s(n,\e/2)$ samples of $\DEC(X)$.
Hence, {\em with probability at least~0.9,
a sequence of $s$ samples of $\DEC(X)$ yields
a collision pattern that leads $T''$ to reject}.
Recall, however, that we invoke $T''$ on $s$ samples
of $X$, not of $\DEC(X)$.
Nevertheless, we show that our tester (i.e.,~$T$)
will reject with high probability also in this case.

\Bcm{\em(The distance of $\DEC(X)$ from $\D$):}
\label{li+sc:reject-clm}
If $\DEC(X)$ is $\e/2$-TV-far from $\D$,
then $T$ rejects with probability $0.9-o(1)$.
\Ecm

\Bpf
We consider $s$ samples $x^{(1)},\dots,x^{(s)}$ taken from $X$.
On the one hand, if any of these $x^{(i)}$'s is $\delta$-H-far from $\Pi$,
then Step~2 rejects with very high probability.
On the other hand, if $x^{(i)}$ is $\delta$-H-close to $\Pi$,
then $\prob[C^{x^{(i)}}(j)\in\{\DEC(x^{(i)})_j,\bot\}]=1-o(1/s^2)$
for every $j\in[n]$,
which means that $T$ either obtains $s$ samples of $\DEC(X)_I$ or rejects.
Recall that, with very high probability,
a sequence of $s$ samples of $\DEC(X)_I$
has the same collision pattern as a sequence of $s$ samples of $\DEC(X)$,
since $\DEC(X)$ is supported by strings that are
pairwise $\delta$-H-far apart.
Lastly, recall that the collision pattern of a sequence of $s$ samples
of $\DEC(X)$ causes $T''$ to reject (whp).
To summarize, letting $C^x(i_1,\dots,i_\ell)=(C^x(i_1),\dots,C^x(i_\ell))$,
we have
\begin{eqnarray*}
\lefteqn{\prob_{x^{(1)},\dots,x^{(s)}\sim X}
         [T^{x^{(1)},\dots,x^{(s)}}(n,\e)\!=\!0]} \\
&\geq& \prob_{x^{(1)},\dots,x^{(s)}\sim X\,;\,I\in{{[n]}\choose\ell}}
               [T''(n,\e;\CP(C^{x^{(1)}}(I),\dots,C^{x^{(s)}}(I))\!=\!0] \\
&\geq& \prob_{x^{(1)},\dots,x^{(s)}\sim X\,;\,I\in{{[n]}\choose\ell}}
            [T''(n,\e;\CP(\DEC(x^{(1)})_I,\dots,\DEC(x^{(s)})_I)\!=\!0]-o(1) \\
&\geq& \prob_{x^{(1)},\dots,x^{(s)}\sim X}
            [T''(n,\e;\CP(\DEC(x^{(1)}),\dots,\DEC(x^{(s)}))\!=\!0]-o(1)
\end{eqnarray*}
which is $0.9-o(1)$.
We stress that the foregoing inequalities hold
since we have ignored cases that cause rejection
(e.g., $x^{(i)}$ being $\delta$-H-far from $\Pi$
and other cases in which $C$ outputs $\bot$).
\Epf

\medskip
Combining Claims~\ref{li+sc:tvd-clm} and~\ref{li+sc:reject-clm},
we infer that if $\DEC(X)$ is $\e/2$-far from $\DP$,
then $T$ rejects~$X$ with high probability.
Recalling that if $X$ is $\e$-far from $\DP$,
then either $X$ is $\e/2$-far from being supported on $\Pi$
(which causes Step~1 to reject (whp))
or $\DEC(X)$ is $\e/2$-far from $\DP$,
it follows that $T$ rejects (whp) in any case.
\EPF


\section{Distributions as Materialization of an Ideal Object}
\label{material-ideal:sec}
As stated in the introduction, we consider three types
of random variations of an ideal object:
random noise applied to bits of a string (a.k.a perturbations),
random cyclic-shifts of a string,
and random isomorphic copies of a graph represented by a string.
These types are studied in the following three subsections.

\subsection{Perturbation}
For two constant parameters $\eta\in[0,0.5)$ and $\delta\in[0,1]$,
and every string $x^*\in\bitset^n$, we consider all distributions
in which each bit of $x^*$ is flipped with probability at most $\eta$
and the outcome is at Hamming distance at most $\delta\cdot n$ from $x^*$.
That is, $\Dper(x^*)$ 
contains the distribution~$X$ if
\begin{enumerate}
\item For every $i\in[n]$, it holds that $\prob[X_i\!\neq\!x^*_i]\leq\eta$.
\item $\prob[|\{i\!\in\![n]\!:\!X_i\!\neq\!x^*_i\}|\leq\delta\cdot n]=1$.
\end{enumerate}
Indeed, setting $\delta=1$ trivializes the second condition,
whereas setting $\delta=0$ mandates $X\equiv x^*$.
Letting $\Dper =\bigcup_{x^*\in\bitset^n}\Dper(x^*)$,
we prove the following.

\BT{\em(Testing noisy versions of a string):}
\label{pertubation:thm}
For two constant parameters $\eta\in[0,0.5)$ and $\delta\in[0,1]$,
the property 
$\Dper$
can be 
tested with $\poly(1/\e)$ queries.
\ET

\BPF
The key observation is that if $X$ is in $\Dper(x^*)$,
for some string $x^*\in\bitset^n$, then each bit of $x^*$ can be
recovered with probability $1-2^{-t}$ by querying $O(t)$ samples
of $X$ (at the corresponding location). This allows to estimate
the flipping probability of individual bits in $X$ as well as
the distribution of the Hamming distance between $X$ and $x^*$.
In view of this observation, the  tester  proceeds as follows
(assuming $\eta+0.25\e<0.5$, or else $\e$ is set so that it satisfies
this constraints (recall that $\eta$ is a constant)).

\begin{enumerate}
\item The tester selects independently and uniformly
at random $\tildeO(1/\e^2)$ indices in $[n]$
and lets the resulting set be denoted by $I$.
\item For each $i\in I$,
the tester estimates the probability $\prob[X_i\!=\!1]$
by taking $\tildeO(1/\e^2)$ samples of~$X$ and querying
each sample at location $i$.
If the estimated value is in $[\eta+0.2\e,1-\eta-0.2\e]$,
then the tester {\em rejects}.
Otherwise it determines $\hat{x}_i$ accordingly;
that is, $\hat{x}_i=0$ if the estimate is at most $\eta+0.2\e$,
and $\hat{x}_i=1$ otherwise.

(Note that the same samples can be used for all $i\in I$.)

\item The tester takes $m=\tildeO(1/\e)$ samples of $X$,
denoted $x^{(1)},\dots,x^{(m)}$.
If, for any $j\in[m]$, it holds that
$|\{i\!\in\!I\!:\!x^{(j)}_i\neq \hat{x}_i\}|>(\delta+0.1\e)\cdot|I|$,
then the tester {\em rejects}.
Otherwise, the tester accepts.
\end{enumerate}
Indeed, for each $i\in I$, the value $\hat{x}_i$
is a candidate for the $i^\xth$ bit of
an ideal object whose perturbation is $X$;
that is, if the tester accepts with high probability,
then it holds that $X$ is in $\Dper(x^*)$
for some $x^*$ such that $x^*_I=\hat{x}_I$.

We first consider the case that $X$ is in $\Dper$.
Specifically, suppose that $X$ belongs to $\Dper(x^*)$
for some $x^*\in \bitset^n$.
First, observe that for any choice of the subset $I$
(in the first step of the algorithm),
the following holds
(by applying the additive Chernoff bound and a union bound):
{\em With high constant probability over the choice of the sampled strings
obtained in the second step, the tester does not reject in this step,
and furthermore, $\hat{x}_i = x^*_i$ for every $i\in I$}.
Next, observe that for any choice of $x^{(1)},\dots,x^{(m)}$
(as obtained in the third step of the algorithm),
the following also holds
(by applying the additive Chernoff bound and a union bound):
{\em The probability over the choice of $I$,
that for some $j$ it holds that
$|\{i\!\in\!I\!:\!x^{(j)}_i\neq x^*_i\}|>(\delta+0.1\e)\cdot|I|$
is a small constant}.
(Note that here we are referring to $x^*$ and not $\hat{x}$).
By combining the two observations we infer that
the tester accepts with high constant probability
(taken both over the choice of $I$
and over the choice of the sample obtained in the second step).

We now consider the case that $X$ is $\e$-far from $\Dper$.
For each $i\in [n]$, let $x'_i$ denote the more likely value of $X_i$;
that is, $\prob[X_i\!=\!x'_i]\geq1/2$.
Then, one of the following two conditions must hold
(or else $X$ is $\e$-close to $\Dper(x')$).
\begin{enumerate}
\item $\sum_{i\in[n]}\min(\prob[X_i\!\neq\!x'_i]-\eta,0)>\e n/2$.

In this case $\prob[X_i\!\neq\!x'_i]>\eta+\e/4$
for at least $\e/4$ fraction of the indices $i\in[n]$.
\item
The probability that $X$ is $(\delta+0.2\e)$-H-far from $x'=x'_1\cdots x'_n$
is at least $0.3\e$.
\end{enumerate}
Suppose that the first condition holds.
Then, with high constant probability over the choice of $I$,
for at least one of the indices $i\in I$,
it holds that $\prob[X_i\!\neq\!x'_i]>\eta+\e/4$,
which implies $\prob[X_i\!=\!x'_i]\in[0.5,1-\eta-0.25\e]$,
whereas $0.5>\eta+0.25\e$.
Assuming this event holds,
{\em with high constant probability over the choice of the sample
obtained in the second step of the algorithm,
the algorithm rejects in this step}.
Next, suppose that the second condition holds.
Then, with high constant probability
(over the sample obtained in the third step),
for at least one of the sample strings $x^{(j)}$
obtained in the third step of the algorithm,
$x^{(j)}$ is $(\delta+0.2\e)$-H-far from $x'$.
Conditioned on this event,
with high constant probability over the choice of $I$,
it holds that $|\{i\!\in\!I\!:\!x^{(j)}_i\neq x'_i\}|>(\delta+0.1\e)\cdot|I|$.
Recall that (for any choice of $I$)
if $I$ contains an index $i$ such that $\prob[X_i\!\neq\!x'_i]>\eta+\e/4$,
then the second step rejects
(with high constant probability over the choice of the sample),
whereas otherwise
(with high constant probability over the choice of the sample)
for every $i\in I$,
the bit $\hat{x}_i$ recovered by the algorithm equals $x'_i$.
In this case, it holds that
$|\{i\!\in\!I\!:\!x^{(j)}_i\neq\hat{x}_i\}|>(\delta+0.1\e)\cdot|I|$,
and {\em the algorithm rejects in the third step}.
Thus, if $X$ is $\e$-far from $\Dper$,
then with high constant probability, the algorithm rejects.
\EPF

\paragraph{Properties of the ideal object.}
For $\eta$ and $\delta$ as above,
and for a property of $n$-bit long strings~$\Pi$,
we let $\DperPi=\bigcup_{x^*\in\Pi}\Dper(x^*)$.
Building on the proof of Theorem~\ref{pertubation:thm},
we get

\BT{\em(Testing noisy versions of a string in a predetermined set):}
\label{noisy-property:thm}
Let $\eta\in[0,0.5)$ and $\delta\in[0,1]$ be constants,
and $\Pi$ be a property of $n$-bit strings that
can be tested using $Q(n,\e)$ queries.
Then, the property $\DperPi$ can be 
tested
using $\poly(1/\e)+\tildeO(Q(n,e/2))$ queries.
\ET

\BPF
We combine the tester presented in the proof
of Theorem~\ref{pertubation:thm} with an emulation
of the tester for $\Pi$. Specifically, each query
made by the latter tester is emulated by making
corresponding queries to $O(\log Q(n,e/2))$ samples
of the tested distribution (and taking a majority vote).
Note that that
both testers are invoked with proximity parameter set to $\e/2$,
and that each emulated query returns the correct answer
(w.r.t the relevant $x^*$)
with probability at least $1-o(1/Q(n,\e/2))$.

Evidently, any distribution $X$ in $\DperPi$
is accepted with high probability.
Likewise, if $X$ is $\e/2$-far from $\Dper$,
then it is rejected with high probability (by the $\Dper$-tester).
Hence, we are left with the case that $X$
is $\e/2$-close to $\Dper(x^*)$ for some $x^*$
that is $\e/2$-H-far from $\Pi$
(because otherwise $X$ is $\e$-close to $\DperPi$).%
\footnote{Specifically, suppose that $X$ is $\e/2$-close
to $\Dper(x^*)$ for some $x^*$
that is $\e/2$-H-close to $z\in\Pi$.
Then, $X$ is $\e$-close to $\Dper(z)$.}
%
Consequently, the emulated tester of $\Pi$
will rejected with high probability.
\EPF

\subsection{Random cyclic shifts}
For any string $x^*\in\bitset^n$, we consider all distributions
that are obtained by random (cyclic) shifts of the string $x^*$;
that is, $\Dcyc(x^*)$ contains the distribution $X$ if there exists
a (related) random variable $J\in\{0,1,\dots,n-1\}$ such that, for every $j$,
with probability $\prob[J\!=\!j]$ it holds that $X_i=x^*_{(i+j)_n}$
for every $i\in[n]$, where $(i+j)_n$ denotes $i+j$ if $i+j\in[n]$
and $i+j-n$ otherwise (i.e., $i+j>n$).

\BT{\em(Testing random shifts of a string):}
\label{random-shifts:thm}
The property $\Dcyc\eqdef\bigcup_{x^*\in\bitset^n}\Dcyc(x^*)$
can be tested using $O(1/\e)$ samples and $\tildeO(\sqrt{n}/\e)$ queries.
\ET
Analogously to Theorem~\ref{noisy-property:thm},
we can also test the ideal string for a predetermined property
provided that this property is invariant under cyclic shifts.
\medskip

\BPF
For the sake of the presentation, we describe a slightly simpler tester
that makes $\tildeO(\sqrt{n}/\e^2)$ queries;
the claimed tester can be obtained by employing
Levin's Economical Work Investment Strategy~\cite[Sec.~8.2.4]{G:pt}.

The straightforward
tester for equality between a pair of $n$-bit long strings
consists of selecting $O(1/\e)$ random indices in $[n]$
and comparing the values of the corresponding bits in the two strings.
To test equality between cyclic versions of the two strings
we try ${\sqrt n}$ random shifts for each string and implicitly
test equality among all resulting ${\sqrt n}\times{\sqrt n}$
by querying each of the resulting $2\cdot{\sqrt n}$ strings
on the same $O(1/\e)$ (or rather $O((\log n)/\e)$) indices.
Since we will want to check ``equality modulo cyclc-shifts''
between $O(1/\e)$ strings, the number of indices (called offsets)
and shifts will be somewhat bigger. Details follow.

The tester is given oracle access to $t=O(1/\e)$ samples of $X$,
denoted $x^{(1)},\dots,x^{(t)}$, and consists of checking
that each $x^{(i)}$ is a cyclic shift of $x^{(1)}$.
Denoting the two strings by $x$ and $y$,
we check whether $y$ is a cyclic shift of $x$ by
selecting $m=O({\sqrt{n\cdot\log t}})$ random shifts,
denoted $s_1,\dots,s_m\in[n]$,
and $\ell=O(\e^{-1}\cdot\log(n/\e))$ random offsets,
denoted $o_1,\dots,o_\ell\in[n]$,
querying both strings at locations $(s_j+o_k)_n$
for every $j\in[m]$ and $k\in[\ell]$,
and accepting if and only if there exists $j,j'\in[m]$
such that $x_{(s_j+o_k)_n}=y_{(s_{j'}+o_k)_n}$ for every $k\in[\ell]$.

We first consider the case that $X$ is in $\Dcyc$;
that is, suppose that $X$ is in $\Dcyc(x^*)$ for some $x^*\in\bitset^n$.
In this case, each of the samples (i.e., $x^{(i)}$)
is a cyclic shift of $x^*$; that is, for each $i\in[t]$,
there exists a shift $\sigma_i$ such that $x^{(i)}_k=x^*_{(k+\sigma_i)_n}$
for every $k\in[n]$.
Hence, for every $i\in\{2,\dots,t\}$
and every pair $j,j'\in[m]$,
with probability at least $1/n$ over the choice of $s_j,s_{j'}\in[n]$,
it holds that $x^{(1)}_{(k+s_j)_n}=x^{(i)}_{(k+s_{j'})_n}$ for every $k\in[n]$
(equiv., $s_j\equiv s_{j'}+\sigma_i\pmod n$).
Since the events that correspond to different pairs of samples
are pairwise independent,
it follows that, for every $i\in\{2,\dots,t\}$,
with probability at least $1-O(n/m^2)$
over the choice of $s_1,\dots,s_{m}\in[n]$,
it holds that $x^{(1)}_{(k+s_j)_n}=x^{(i)}_{(k+s_{j'})_n}$
for some $j,{j'}\in[m]$ and every $k\in[n]$.
We conclude that, in this case
(regardless of the choice of the $x^{(i)}$'s and the $o_k$'s),
the tester accepts with probability at least~$2/3$.

Suppose, on the other hand, that $X$ is $\e$-far from $\Dcyc$.
Fixing the first sample, denoted $x^{(1)}$,
it follows that with probability at least $\e/2$
it holds that (a sample of) $X$ is $(\e/2)$-H-far from
being some shift of $x^{(1)}$.
Hence, with probability at least 0.9 over the choice of the $x^{(i)}$'s,
there exists an $i\in[t]$ such that $x^{(i)}$ is $(\e/2)$-H-far from
being any shift of $x^{(1)}$.
It follows that, for each choice of $s_1,\dots,s_{m}\in[n]$
and every $j,j'\in[m]$, it holds that
$|\{k\!\in\![n]:x^{(1)}_{(k+s_j)_n}\neq x^{(i)}_{(k+s_{j'})_n}\}|>\e n/2$,
and consequently for every $j,j'\in[m]$ it holds that
$$\prob_{o_1,\dots,o_\ell\in[n]}
   \left[(\exists k\in[\ell])\;
         x^{(1)}_{(s_j+o_k)_n}\neq x^{(i)}_{(s_{j'}+o_k)_n}\}|
    \right] \;>\; 1-(1-0.5\e)^\ell.$$
It follows that with probability
at least $1-m^2\cdot\exp(-\e\cdot\ell)>0.9$
(over the choice of $o_1,\dots,o_k$)
the tester detects that $x^{(i)}$ is not a cyclic shift of $x^{(1)}$,
where the inequality uses $m=O({\sqrt{n\cdot\log t}})$
and a sufficiently large $\ell=O(\e^{-1}\cdot\log(n/\e))$.
Therefore, in this case (i.e., $X$ $\e$-far from $\Dcyc$),
the tester rejects with probability at least~$2/3$.
This completes the analysis of the slightly simpler tester,
which performs $t\cdot m\cdot\ell=\tildeO(\sqrt{n}/\e^2)$ queries.

The claimed tester (which performs $\tildeO(\sqrt{n}/\e)$ queries)
follows by observing that if $X$ is $\e$-far from $\Dcyc$,
then, for some $r\in[\log(2/\e)]$,
with probability at least $2^{r}\cdot\e/O(\log(1/\e))$
it holds that (a sample of) $X$ is $2^{-r}$-H-far from
being a shift of $x^{(1)}$.
Hence, it suffices to have $O(\log(1/\e))$ iterations
such that in the $r$-th iteration we use $t=2^{-r}\cdot\tildeO(1/\e)$
and $\ell=O(2^{r}\cdot\log(n/\e))$.
\EPF

\paragraph{Testing specific distributions over cyclic shifts.}
The property $\Dcyc$ does not impose any constraint
on the distribution over shifts.
We next consider a natural variant, where this distribution is uniform.

\BT{\em(Testing uniformly random shifts of a string):}
\label{uniform-shifts:thm}
Let $\Dcycu(x^*)$ denote the uniform distribution over
the cyclic shifts of a string $x^*$.
Then, the property $\Dcycu\eqdef\bigcup_{x^*\in\bitset^n}\Dcycu(x^*)$ can
be tested using $\tildeO(n^{2/3}/\e^3)$ queries.
\ET
Theorem~\ref{uniform-shifts:thm} is proved by a reduction
to a more general problem, and it is indeed possible
that a more efficient tester exists.
\medskip

\BPF
We reduce the current problem to testing the equality between
two distributions over $\bitset^n$
such that one of the distributions has support size at most $n$,
while noting that a tester for the latter problem is provided
in Theorem~\ref{equal-distributions:thm}.
Specifically, given $s$ samples, denoted $x^{(1)},\dots,x^{(s)}$,
of a distribution $X$ over $n$-bit strings,
we consider the distribution $Y\eqdef\Dcycu({x^{(1)}})$,
and test equality between $X$ and $Y$,
where we emulate samples to $X$ by using $x^{(2)},\dots,x^{(s)}$,
and emulate samples to $Y$ by using (random shifts of) $x^{(1)}$.
Note that $Y$ has support of size at most $n$,
which suffices when using the furthermore clause
of Theorem~\ref{equal-distributions:thm}.

The complexity of our tester equals the complexity
of the tester of Theorem~\ref{equal-distributions:thm},
and its analysis reduces to the latter.
Specifically,  if $X$ is in $\Dcycu$,
then, for every possible $x^{(1)}$ drawn from $X$,
it holds that $X\equiv\Dcycu(x^{(1)})$,
and it follows that our tester accepts (whp).
On the other hand, if $X$ is $\e$-far from $\Dcycu$,
then for every $x^*$ it holds that $X$ is $\e$-far from $\Dcycu(x^*)$,
and it follows that our tester rejects (whp).
\EPF

\BR{\rm(Generalization of Theorem~\ref{uniform-shifts:thm}):}
\label{uniform-shifts:rem}
The proof of Theorem~\ref{uniform-shifts:thm} generalizes
to any fixed distribution $I$ over $\{0,1,...,n-1\}$
such that for every $i$ in the support of $I$
it holds that $I\equiv I+i\pmod n$.
Specifically, let $\DcycR(x^*)$ denote the distribution obtained
by shifting $x^*$ by~$I$ positions {\rm(to the right)}.
Then, the property $\DcycR\eqdef\bigcup_{x^*\in\bitset^n}\DcycR(x^*)$ can
be tested using $\tildeO(n^{2/3}/\e^3)$ queries.
The generalization follows by observing that
for any $x^{(1)}$ in the support of $\DcycR(x^*)$
it holds that $\DcycR(x^{(1)})\equiv\DcycR(x^*)$.
\ER

\subsection{Random isomorphic copies of a graph}
Using a sublinear-query tester for graph isomorphism,
we can adapt the ideas underlying the proof of
Theorem~\ref{random-shifts:thm} to test distributions
of strings that describe the adjacency matrices of
random isomorphic copies of a graph.
That is, we consider $n$-bit long strings that describe
the adjacency matrices of $\sqrt n$-vertex graphs.
Specifically, for every string $x^*\in\bitset^n$,
we consider the graph $G_{x^*}$ described by $x^*$
and any distribution on isomorphic copies of $G_{x^*}$;
that, $\Diso(x^*)$ contains the distribution $X$ if $X$
is a distribution over strings that describe graphs
that are isomorphic to $G_{x^*}$.

Recall that testing isomorphism of $k$-vertex graphs
in the dense graph model,
which uses the adjacency matrix representation,
has query complexity $\poly(1/\e)\cdot\tildeO(k^{5/4})$;
see~\cite{FM},
where the dependence on $\e$ is mentioned at the end of Section~1.
In contrast, the query complexity of the 
tester of~\cite{OS}
is $k^{1+o(1)}$ provided that $\e=\omega((\log\log k)/(\log k)^{1/2})$.
(For an introduction to the dense graph model see~\cite[Chap.~8]{G:pt}.)

\BT{\em(Testing random isomorphic copies of a graph):}
\label{random-isomorphic-copies:thm}
\sloppy
The property $\Diso\eqdef\bigcup_{x^*\in\bitset^n}\Diso(x^*)$
can be tested using $O(1/\e)$ samples
and $\poly(1/\e)\cdot\tildeO(n^{5/8})$ queries.
\ET
Note that testing isomorphism in the dense graph model is
reducible to testing $\Diso$ in the \DHOm.
We also mention that, analogously to Theorem~\ref{noisy-property:thm},
one can also test the ideal string for a predetermined graph property
(since a graph property is invariant under graph isomorphism).
\medskip

\BPF
Analogously to the proof of Theorem~\ref{random-shifts:thm},
the tester takes $t=O(1/\e)$ samples, denoted $x^{(1)},\dots,x^{(t)}$,
and checks whether all $x^{(i)}$'s describe graphs that are
isomorphic to the graph described by $x^{(1)}$.
Hence, for each $i\in\{2,\dots,t\}$,
we check whether $G_{x^{(i)}}$ is isomorphic to $G_{x^{(1)}}$,
by invoking a graph isomorphism tester for the dense graph model.
Specifically, we use the tester presented in~\cite{FM},
while setting the proximity parameter to $\e/2$
(and the error probability of the test to $o(\e)$).

Note that if $X$ is in $\Diso$,
then each invocation of the isomorphism test accept
with probability $1-o(\e)$.
On the other hand, if $X$ is $\e$-far from $\Diso$
(according to Definition~\ref{dist:def}),
then, for any choice of $x^{(1)}$ and every $i\in\{2,\dots,t\}$,
with probability at least $\e/2$ it holds that $G_{x^{(i)}}$
is $\e/2$-far from being isomorphic to $G_{x^{(1)}}$,
where the latter distance is in the dense graph model.
Hence, with high constant probability, there exists $i\in\{2,\dots,t\}$
such that $G_{x^{(i)}}$ is $\e/2$-far from being isomorphic to $G_{x^{(1)}}$,
and the corresponding invocation of the graph isomorphism tester
rejects (w.h.p).
\EPF

\paragraph{What about the bounded-degree graph model?}
We could have adapted the proof strategy
of Theorem~\ref{random-isomorphic-copies:thm}
to bounded-degree graphs that are represented by their incidence functions
(see~\cite[Chap.~9]{G:pt} for introduction to this model).
However, unfortunately, we do not know of a sublinear-query tester
for graph isomorphism in that model.
Instead, one can prove results analogous to Theorem~\ref{noisy-property:thm}
with respect to graph properties that have a sub-linear isomorphism test
as well as a sub-linear tester (see, e.g.,~\cite{G:iso}).


\section{Tuples of Distributions}
\label{tuples:sec}
Our notion of testing properties of distributions over huge objects
(as captured by Definition~\ref{dist:test-one.def}),
extends easily to testing tuples of such distributions.

\subsection{The definition}
Following the convention stated in Section~\ref{conventions:sec},
we refer to distributions via the corresponding random variables.

\BD{\rm(Testing properties of $t$-tuples of huge distributions):}
\label{dist:test-two.def}
For $t\in\N$, let $\cal D$ be a property of $t$-tuples of distributions,
which are each as in Definition~\ref{dist:test-one.def},
and $s:\N\times(0,1]\to\N$.
A {\sf tester}, denoted $T$, {\em of sample complexity $s$}
{\sf for the property $\cal D$} is a probabilistic machine that,
on input parameters $n$ and $\e$,
and oracle access to a sequence of $s(n,\e)$ samples drawn from
each of the $t$ unknown distributions $X^{(1)},\dots,X^{(t)}\in\bitset^n$,
satisfies the following two conditions.
\begin{enumerate}
\item{\em The tester accepts tuples that belong to $\cal D$:}
If $(X^{(1)},\dots,X^{(t)})$ is in $\cal D$, then
$$\prob_{x^{(1,1)},\dots,x^{(1,s)}\sim X^{(1)}
  ;\ldots;x^{(t,1)},\dots,x^{(t,s)}\sim X^{(t)}}
  [T^{x^{(1,1)},\dots,x^{(1,s)},\dots,x^{(t,1)},\dots,x^{(t,s)}}(n,\e)\!=\!1]
  \geq2/3,$$
where $s=s(n,\e)$
and $x^{(i,1)},\dots,x^{(i,s)}$ are drawn independently from
the distribution $X^{(i)}$.
\item{\em The tester rejects tuples that are far from $\cal D$:}
If $(X^{(1)},\dots,X^{(t)})$ is $\e$-far from $\cal D$
{\em(i.e., for every $(Y_1,\dots,Y_t)$ in $\cal D$ the average distance
(according to Definition~\ref{dist:def})
between $X_j$ and~$Y_j$, where $j\in[t]$, is greater than $\e$)},
then
$$\prob_{x^{(1,1)},\dots,x^{(1,s)}\sim X^{(1)}
  ;\ldots;x^{(t,1)},\dots,x^{(t,s)}\sim X^{(t)}}
  [T^{x^{(1,1)},\dots,x^{(1,s)},\dots,x^{(t,1)},\dots,x^{(t,s)}}(n,\e)\!=\!0]
  \geq2/3,$$
where $s=s(n,\e)$ and $x^{(i,1)},\dots,x^{(i,s)}$
are as in the previous item.
\end{enumerate}
\ED
The query complexity of such tester is defined as
in the case of testing a single distribution (i.e., $t=1$).
Indeed, Definition~\ref{dist:test-one.def} is a special case
of Definition~\ref{dist:test-two.def} (i.e., $t=1$).

\subsection{Testing equality}
Testing equality of two distributions
is the archetypal example for the case of $t=2$.
Using any tester for the standard model, we obtain a tester
for the \DHO\ model by querying each sample at
a logarithmic (in the support size) number of locations.

\BT{\em(Testing equality of distributions in the \DHO\ model):}
\label{equal-distributions:thm}
For any $m,n\in\N$ and $\e>0$,
given a pair of distributions over $\bitset^n$
that have support size at most $m$,
we can distinguish between the case that they are identical
and the case that they are $\e$-far from one another
{\em(according to Definition~\ref{dist:def})}
using $\tildeO(m^{2/3}/\e^3)$ queries and $O(m^{2/3}/\e^2)$ samples.
Furthermore, the claim holds even if only the support size
of one of the distributions is upper-bounded by~$m$.
\ET
Actually, the sample complexity
is $s=O(\max(\e^{-4/3} m^{2/3}, \e^{-2} m^{1/2}))$,
and the query complexity is~$\tildeO(s/\e)$.
\medskip

\BPF
The key observation is that if $X$ is $\e$-far from $Y$
(according to Definition~\ref{dist:def}), then,
with high probability over the choice of
a random $O(\e^{-1}\log m)$-subset $J\subset[n]$,
the total variation distance between $X_J$ and $Y_J$
is at least $0.3\e$.
This observation is proved next. 

\Bcm{\em(Typically, the distance between $X$ and $Y$
is preserved by the distance between~$X_J$ and~$Y_J$):}
\label{two-distributions:clm}
Suppose that $X$ is $\e$-far from $Y$,
and that both distributions have support size at most~$m$.
Then, with probability $1-o(1)$ over the choice
of $J\in{{[n]}\choose{O(\e^{-1}\log m)}}$,
the total variation distance between $X_J$ and $Y_J$
is greater than $0.3\e$.
Actually, $X_J$ is $0.3\e$-far from $Y_J$,
even according to Definition~\ref{dist:def}.
\Ecm

\Bpf
We start by letting $x^{(1)},\dots,x^{(m')}$
(resp., $y^{(1)},\dots,y^{(m'')}$)
denote the elements in the support of $X$ (resp., $Y$),
where $m'\leq m$ (resp., $m''\leq m$).
Next, we note that for every $i\in[m']$ and $k\in[m'']$,
when selecting uniformly an $O(t/\e)$-subset $J$,
with probability at least $1-2^{-t}$,
the relative Hamming distance between $x^{(i)}_J$ and $y^{(k)}_J$
it at least
$\min(0.5 \cdot \Delta_H(x^{(i)},y^{(k)}),\Delta_H(x^{(i)},y^{(k)})-0.2\e)$,
where the first (resp., second) term refers to
the case that $\Delta_H(x^{(i)},y^{(k)})>0.2\e$
(resp., $\Delta_H(x^{(i)},y^{(k)})\leq0.2\e$)
and followed by a multiplicative Chernoff bound (resp., by triviality).
%

Now, letting $t=O(\log m)$
and using a union bound (over all $(i,k)\in[m']\times[m'']$),
with probability $1-o(1)$ over the choice of $J\in{{[n]}\choose{O(t/\e)}}$,
for every $i\in[m']$ and every mapping $\mu:[m']\to[m'']$,
it holds that
\begin{equation} \label{equal-distributions:eq0}
 \Delta_H(x^{(i)}_J,y^{(\mu(i))}_J)
 \geq 0.5\cdot\Delta_H(x^{(i)},y^{(\mu(i))}) -0.2\e.
\end{equation}
(We stress the order of quantifiers: With high probability
over the choice of~$J$, \eqref{equal-distributions:eq0}
holds for every $i\in[m']$ and $\mu:[m']\to[m'']$.)%
\footnote{Indeed, although there are $(m'')^{m'}$ different $\mu$'s,
the union bounds is only over the pairs $(i,\mu(i))\in[m']\times[m'']$.}
Hence, with probability $1-o(1)$ over the choice of~$J$,
for every mapping $\mu:[m']\to[m'']$
and every probability distribution $p:[m']\to[0,1]$,
it holds that
\begin{eqnarray}
\label{equal-distributions:eq0a}
\sum_{i\in[m']}p(i)\cdot\Delta_H(x^{(i)}_J,y^{(\mu(i))}_J)
&\geq& \sum_{i\in[m']}p(i)\cdot
                (0.5\cdot\Delta_H(x^{(i)},y^{(\mu(i))})-0.2\e) \\
&\geq& 0.5\cdot\sum_{i\in[m']}
               p(i)\cdot\Delta_H(x^{(i)},y^{(\mu(i))})\;\;-\;0.2\e.
\label{equal-distributions:eq0b}
\end{eqnarray}
Suppose now that instead of a (deterministic) mapping $\mu:[m']\to[m'']$,
we consider a randomized process $\mu:[m']\to[m'']$,
where for each $i\in [m']$ the random variable $\mu(i)$
represents a distribution over $[m'']$.
Then, \eqrefs{equal-distributions:eq0a}{equal-distributions:eq0b}
extends to any random process $\mu$,
where we consider expected distances
(with expectation taken over the random choices of $\mu$).
In particular, letting $p(i)=\prob[X\!=\!x^{(i)}]$,
with probability $1-o(1)$ over the choice of $J$,
for any randomized process $\mu:[m']\to[m'']$, it holds that
\begin{equation} \label{equal-distributions:eq1}
\sum_{i\in[m']}p(i)
   \cdot \Exp_\mu\left[\Delta_H(x^{(i)}_J,y^{(\mu(i))}_J)\right]
\geq 0.5\cdot \sum_{i\in[m']}p(i)
      \cdot \Exp_\mu\left[\Delta_H(x^{(i)},y^{(\mu(i))})\right]
      \;-0.2\e.
\end{equation}
We observe that for any choice of $\mu$ that maps $X$ to $Y$
(i.e., $\sum_{i\in[m']}p(i)\cdot\prob_\mu[\mu(i)\!=\!k]=\prob[Y\!=\!y^{(k)}]$
for every $k\in[m'']$),
the main sum in the r.h.s of \eqref{equal-distributions:eq1}
is lower-bounded by the distance between $X$ and $Y$
(according to Definition~\ref{dist:def}; cf.~\eqref{general:cost-of-move:eq}).
Recalling that the latter distance is greater than $\e$,
it follows that (for any $\mu$ that maps $X$ to $Y$)
the l.h.s of \eqref{equal-distributions:eq1}
is greater than $0.5\cdot\e-0.2\e=0.3\e$.
On the other hand, we observe that the minimum
of the l.h.s of \eqref{equal-distributions:eq1}
taken over $\mu$'s that map $X_J$ to $Y_J$
captures the distance between $X_J$ and $Y_J$
(according to Definition~\ref{dist:def}).
This establishes the furthermore claim
(i.e., with probability $1-o(1)$ over the choice of $J$,
the distance between $X_J$ and $Y_J$ is greater than~$0.3\e$).
The main claim follows
(since the distance according to Definition~\ref{dist:def}
lower-bounds the total variation distance).
\Epf

\myparagraph{The tester.}
In light of the above, our tester proceeds as follows.
For $s=O(\max(\e^{-4/3} m^{2/3},\allowbreak \e^{-2} m^{1/2}))$,
given oracle access to $s$ samples,
denoted $u^{(1)},\dots,u^{(s)}$ and $v^{(1)},\dots,v^{(s)}$,
of each of the two distributions, the tester selects
an $O(\e^{-1}\log m)$-subset $J\subset[n]$ uniformly at random,
and queries each sample at the bits in $J$.
Denoting the 
resulting strings
(i.e., the restrictions of the sampled strings to $J$)
by $u^{(1)}_J,\dots,u^{(s)}_J$ and $v^{(1)}_J,\dots,v^{(s)}_J$,
our tester invokes the standard tester (with proximity parameter $0.3\e$),
and provides these strings as the expected samples.

Note that if $X\equiv Y$, then $X_J\equiv Y_J$ always holds,
and the standard tester accepts (whp).
On the other hand, by the foregoing observation,
if $X$ is $\e$-far from $Y$
(according to Definition~\ref{dist:def}), then,
with high probability over the choice of $J$,
it holds that $X_J$ is $0.3\e$-far from $Y_J$
(in total variation distance), and in this case
the standard tester rejects (whp).

\myparagraph{Proving the furthermore claim of the theorem.}
Having established the main claim,
we turn to proving the furthermore claim,
where we only assume that $Y$ (rather than also $X$)
has support size at most $m$.
In this case we cannot afford a union bound over $[m']\times[m'']$.
Still, letting $t=O(\log(m/\e))$ and assuming only $m''\leq m$,
we replace the assertion regarding \eqref{equal-distributions:eq0}
by the assertion that, for every $i\in[m']$,
with probability $1-o(\e)$ over the choice of $J\in{{[n]}\choose{O(t/\e)}}$,
for every mapping $\mu:[m']\to[m'']$ it holds that
\begin{equation} \label{equal-distributions:eq0r}
 \Delta_H(x^{(i)}_J,y^{(\mu(i))}_J)
 \geq 0.5\cdot\Delta_H(x^{(i)},y^{(\mu(i))}) -0.2\e.
\end{equation}
Fixing any probability distribution $p:[m']\to[0,1]$,
we call $J$ {\sf good} if for every mapping $\mu:[m']\to[m'']$
it holds that \eqref{equal-distributions:eq0r} is satisfies
for a set of $i$'s that has weight at least $1-0.1\e$ under $p$
(i.e., letting $I$ denote the set of these $i$'s,
it holds that $\sum_{i\in I}p(i)\geq1-0.1\e$).
Using an averaging argument, it follows that $1-o(1)$
of the $J$'s are good.
Hence, with probability $1-o(1)$ over the choice of $J$,
for every $\mu:[m']\to[m'']$ it holds that
\begin{eqnarray}
\label{equal-distributions:eq3a}
\sum_{i\in[m']}p(i)\cdot\Delta_H(x^{(i)}_J,y^{(\mu(i))}_J)
&\geq& \sum_{i\in[m']}p(i)\cdot
                (0.5\cdot\Delta_H(x^{(i)},y^{(\mu(i))})-0.2\e)\;\;-\;0.05\e \\
&\geq& 0.5\cdot\sum_{i\in[m']}
               p(i)\cdot\Delta_H(x^{(i)},y^{(\mu(i))})\;\;-\;0.25\e,
\label{equal-distributions:eq3b}
\end{eqnarray}
where the term $0.05\e$ accounts for contribution of the $i$'s
that do not satisfy \eqref{equal-distributions:eq0r}.
That is, \eqrefs{equal-distributions:eq0a}{equal-distributions:eq0b}
is replaced by \eqrefs{equal-distributions:eq3a}{equal-distributions:eq3b}.
Proceeding as in the proof of the Claim~\ref{two-distributions:clm},
we get

\Bcm{\em(Claim~\ref{two-distributions:clm}, extended):}
\label{two-distributions:clm2}
Suppose that $X$ is $\e$-far from $Y$,
and that $Y$ has support size at most $m$.
Then, with probability $1-o(1)$ over the choice
of $J\in{{[n]}\choose{O(\e^{-1}\log m)}}$,
the total variation distance between $X_J$ and $Y_J$
is greater than $0.25\e$.
Actually, $X_J$ is $0.25\e$-far from $Y_J$,
even according to Definition~\ref{dist:def}.
\Ecm
The furthermore claim (of the theorem)
follows by observing that the equality tester
(for the standard model) of~\cite{DK}
works also when the support size of only one of
the tested distributions is upper-bounded.%
\footnote{This is not a generic claim regarding any such tester:
See Footnote~\ref{std-equality:fn} in Appendix~\ref{std-equality:apdx}.}
%
Specifically, using the presentation of~\cite[Sec.~11.2--11.3]{G:pt},
we observe that the support size is only used in the proof
of~\cite[Cor.~11.21]{G:pt}, when upper-bounding the total variation
distance between two distributions by the norm-2 of their difference.
But essentially the same upper bound (on the total variation
distance) holds also if only the support
of one of the distributions is upper-bounded.%
\footnote{Specifically, let $p:S\to[0,1]$ be the probability function
representing one distribution and $q:U\to[0,1]$ be the function
representing the other distribution, where $S\subseteq U$. Then,
\begin{eqnarray*}
\sum_{i\in U}|p(i)-q(i)|
&=& 2\cdot\sum_{i\in U:p(i)>q(i)}|p(i)-q(i)| \\
&\leq& 2\cdot\sum_{i\in S}|p(i)-q(i)| \\
&\leq& 2\cdot\sqrt{|S|}\cdot\left(\sum_{i\in S}|p(i)-q(i)|^2\right)^{1/2} \\
&\leq& 2\cdot\sqrt{|S|}\cdot\left(\sum_{i\in U}|p(i)-q(i)|^2\right)^{1/2}
\end{eqnarray*}
which equals $2\sqrt{|S|}\cdot\|p-q\|_2$.}
(For more details, see Appendix~\ref{std-equality:apdx}.)
\EPF

\nocite{BFFKRW}

\section*{Acknowledgements}
\addcontentsline{toc}{section}{Acknowledgements}
We are grateful to Avi Wigderson for a discussion
that started this research project.
We also wish to thank the anonymous reviewers
for their comments and suggestions.


\addcontentsline{toc}{section}{References}

\newpage
\printbibliography

\appendix
\setcounter{section}{1}
\setcounter{subsection}{0}
\setcounter{theorem}{0}
\section*{Appendices}
\phantomsection
\addcontentsline{toc}{section}{Appendices}
The appendices vary in nature.
Appendix~\ref{emd-tvd:apdx} presents a proof of a well-known fact,
Appendix~\ref{general-equality:apdx} presents ramifications
on a simple result stated in the main text,
Appendix~\ref{closed-mapping:apdx} presents a possible avenue
towards a stronger version of Theorem~\ref{closed-mapping:ithm},
and Appendix~\ref{std-equality:apdx} presents a result that
is only implicit in prior work (and may be of independent interest).

\subsection{Earth mover distance with inequality measure}
\label{emd-tvd:apdx}
A {\em general definition of the earth model distance}\/
associates a distance function $f$ with the domain,
and considers the cost of the best randomized process
that transforms one distribution to another,
where the cost of the (randomized) process $M$,
which moves the distribution $X$ to the distribution $M(X)$,
is defined as
\begin{equation}\label{general:cost-of-move:eq}
\sum_{x}\prob[X\!=\!x] \cdot\Exp[f(x,M(x))].
\end{equation}
Definition~\ref{dist:def} is derived by letting $f$ be
the relative Hamming distance between strings.%
\footnote{Here the random process $M$ replaces the $w_{x,y}$'s
that appear in \eqref{EMD-via-LP:eq};
specifically, $\prob[M(x)\!=\!y]=\frac{w_{x,y}}{\prob[X\!=\!x]}$.}
Here, we consider the crude inequality function;
that is, $f(x,y)=1$ if $x\neq y$ and $f(x,x)=0$.
We prove that the earth mover's distance with respect to
the inequality function equals the total variation distance,
where the total variation distance between $X$ and $Y$
equals $\max_S\{\prob[X\!\in\!S]-\prob[Y\!\in\!S]\}$,
which equals half $\sum_z|\prob[X\!=\!z]-\prob[Y\!=\!z]|$.

\BCM{\em(On the earth mover's distance with the inequality measure):}
The earth mover's distance with respect to the inequality measure
{\rm(i.e., $f(x,y)=1$ if $x\neq y$ and $f(x,x)=0$)}
coincides with the total variation distance.
\ECM

\BPF
For $S=\{z:\prob[X\!=\!z]\leq\prob[Y\!=\!z]\}$,
consider the randomized process $M$ such that $M(z)=z$ if $z\in S$
and $\prob[M(z)\!=\!z]=\frac{\prob[Y=z]}{\prob[X=z]}$ otherwise
(i.e., $x\in{\ov S}$),
where the excess probability mass (of ${\ov S}$)
is distributed among the strings in $S$ so that $M(X)\equiv Y$.
Note that the cost of this process $M$ equals
$$\sum_{x\not\in S}\prob[X\!=\!x]
    \cdot\left(1-\frac{\prob[Y\!=\!z]}{\prob[X\!=\!z]}\right)$$
which equals the total variation distance.
Hence, the earth mover distance (w.r.t inequality) is upper-bounded
by the total variation distance.
On the other hand,  the earth mover distance (w.r.t inequality)
is lower-bounded by the total variation distance,
since the latter measures the probability mass that has
to be moved from $S=\{z:\prob[Y\!=\!z]\geq\prob[X\!=\!z]\}$
to ${\ov S}$.
\EPF

\subsection{Ramifications regarding Theorem~\ref{support:ithm}}
\label{general-equality:apdx}
We restate the claim of Theorem~\ref{support:ithm}
and improve upon it in the special case of ``nice'' query complexity bounds.
Specifically, we prove the following.

\BT{\em(Theorem~\ref{support:ithm}, restated and improved):}
\label{support:thm}
For a property of bit strings $\Pi=\{\Pi_n\}_{n\in\N}$
such that $\Pi_n\subseteq\bitset^n$,
let $\D_\Pi=\{{\D}_n\}_{n\in\N}$ such that $\D_n$ denotes the set
of all distributions that have a support that is subset of $\Pi_n$.
If the query complexity of testing $\Pi$ is $q$,
then the query complexity of testing $\D_\Pi$ is at most $q'$
such that $q'(n,\e)=\tildeO(1/\e)\cdot q(n,\e/2)$.
Furthermore, if $q(n,\e)\geq2^c\cdot q(n,2\e)$ holds
for some constant $c\geq1$ and all $\e\in[O(1/n),\Omega(1)]$,
then $q'(n,\e)=O(q(n,\e))$ if $c>1$
and $q'(n,\e)=O(\log(1/\e))^3\cdot q(n,\e)$ otherwise {\rm(i.e., if $c=1$)}.
In both cases, the tester uses $O(1/\e)$ samples.
\ET

\BPF
Recall that the proof of the main claim relied
on the observation that if the tested distribution $\cP$
is $\e$-far from $\D_n$ (according to Definition~\ref{dist:def}),
then, $x\sim \cP$ is $\e/2$-H-far from $\Pi_n$
with probability at least $\e/2$.
(This is the case, since otherwise, letting $f(x)$ be a string
in $\Pi_n$ that is closest to $x$ in Hamming distance
yields a distribution $\cQ(y)= \sum_{x\in f^{-1}(y)}\cP(x)$
that is in $\D_\Pi$ and is
$(\frac\e2\cdot1+(1-\frac\e2)\cdot\frac\e2)$-close to $\cP$.)

The furthermore claim is proved by
employing Levin's Economical Work Investment
Strategy~\cite[Sec.~8.2.4]{G:pt}.
Specifically, the key observation is that
there exists $i\in[\ceil{\log_2(16/\e)}]$
such that with probability at least $2^{-i}/(i+1)^2$
it holds that $x\sim P$ is $2^{i-3}\cdot\e$-H-far from $\Pi_n$.
In this case, the query complexity
is $\sum_{i\leq\ell}O(i^2\cdot2^i) \cdot q(n,2^{i-3}\e)$,
where $\ell=\ceil{\log_2(16/\e)}$.
Using $q(n,2^{i-3}\e)\leq(2^{i-3})^{-c}\cdot q(n,\e)$,
the foregoing sum is upper-bounded by
is $\sum_{i\leq\ell}i^2\cdot2^{-(c-1)\cdot i}\cdot O(q(n,\e))$,
and the claim follows.
%
\EPF

\paragraph{Generalization.}
Towards the following generalization of Theorem~\ref{support:thm},
we consider a generalization of property testing of strings.
In this generalization
the property $\Pi_n$ is partitioned into $m(n)$ parts
and, when accepting, the tester also indicates the
index of the part in which the object resides.
For example, the set of low-degree multi-variate polynomials
can be partitioned according to their value at a fixed point,
and coupled with a generalized tester of low complexity.
Generalizing the main claim of Theorem~\ref{support:thm}, we get~--

\BT{\em(Theorem~\ref{support:ithm}, generalized):}
\label{partitioned-support:thm}
Let $\Pi=\{\Pi_n\}_{n\in\N}$ be as in Theorem~\ref{support:thm}.
For $m:\N\to\N$,
suppose that $\Pi_n=\bigcup_{i\in[m(n)]}\Pi_n^{(i)}$
is testable in the generalized sense within query complexity $q(n,\e)$.
Let  ${\cal D}=\{{\cal D}_n\}_{n\in\N}$ be a property
of distributions over $[m(n)]$ that is testable in the
standard model with sample complexity $s(n,\e)=\Omega(1/\e)$,
and let $\C_n=\C_{\Pi_n,\D_n}$ be a property of distributions
such that $X$ is in $\C_n$ if and only if $X$ consists
of selecting an index $i\in[m(n)]$ according
to some distribution in $\D_n$ and outputting an element
selected according to an arbitrary distribution
that is supported by a subset of $\Pi_n^{(i)}$.
Then, the query complexity of testing $\C=\bigcup_{n\in\N}\C_n$
is at most $q'$ such that $q'(n,\e)=\tildeO(s(n,0.3\e))\cdot q(n,0.3\e)$.
\ET
Note that Theorem~\ref{partitioned-support:thm}
generalizes the main claim of Theorem~\ref{support:thm},
but not its furthermore claim.
We comment that
if each $\Pi_n^{(i)}$ is testable with $q(n,\e)$ queries
and the $\Pi_n^{(i)}$'s are $\delta$-H-far apart,
then we can obtain a generalized tester of query complexity
$\tildeO(m(n))\cdot q(n,\delta)+O(q(n,\e))$ for $\Pi_n$;
specifically, we first test each $\Pi_n^{(i)}$
with proximity parameter $\delta$,
while relying on the fact that $x\in\Pi_n^{(i)}$
implies that $x$ is $\delta$-H-far from each of
the other $\Pi_n^{(j)}$'s.
\medskip

\BPF
We combine the tester for $\Pi$, denoted $T$, with the tester for $\D$,
while invoking both with proximity parameter $\e/2$,
and reducing the error probability of $T$ to $o(1/s(n,0.3\e))$.
Hence, when invoked on input $(n,0.3\e)$ and given oracle access
to $x\in\bitset^n$, with probability at least $1-o(1/s(n,0.3\e))$,
the tester $T$ outputs~$i$ if $x\in\Pi_n^{(i)}$,
and rejects (with output~0) if $x$ is $0.3\e$-H-far from $\Pi_n$.
Furthermore, with probability at least $1-o(1/s(n,0.3\e))$,
the tester $T$ does not output~$i$ if $x$ is $0.3\e$-H-far from $\Pi_n^{(i)}$.
Denoting the (majority) output of $T$ by $\chi(x)$,
we may assume that either $\chi(x)=0$ (indicating rejection)
or $x$ is $\e/2$-H-close to $\Pi_n^{(\chi(x))}$.
The key observation is that if~$X$ is $\e$-far from $\C_n$
(according to Definition~\ref{dist:def}),
then either $X$ is $0.7\e$-far from being distributed over $\Pi_n$
(according to Definition~\ref{dist:def})
or $\chi(X)$ is $0.3\e$-TV-far from $\D_n$.
Hence, we get an adequate tester that, on access
to the samples $x^{(1)},\dots,x^{(s)}$, where $s=s(n,0.3\e)$,
invokes $T$ on each of these samples,
obtaining the answers $a_1,\dots,a_s\in\{0,1,\dots,m(n)\}$,
rejects if any of these $a_i$'s equals~0,
and outputs the verdict of the distribution tester (i.e., the $\D$-tester)
on $(a_1,\dots,a_s)\in[m(n)]^s$ otherwise.

To see that the foregoing tester is correct,
note that if $X$ in $\C_n$,
then $X=Y_I$ such that $I$ is in $\D_n$
and each $Y_i$ is supported by $\Pi_n^{(i)}$.
It follows that, in this case, $X$ is accepted with high probability.
On the other hand, if $X$ is accepted with high probability,
then $\chi(X)$ is $0.3\e$-TV-close to a distribution in $\D_n$,
and, with probability at least $1-0.3\e$ over the choice of $x\sim X$,
it holds that $x$ is $0.3\e$-H-close to $\Pi_n^{(\chi(x))}$.
It follows that $X$ is $\e$-close to $\C_n$.
\EPF

\subsection{Towards a stronger version of Theorem~\ref{closed-mapping:ithm}}
\label{closed-mapping:apdx}
Recall that, for any property $\D$ that is closed under mapping,
Theorem~\ref{closed-mapping:ithm} upper-bounds
the query complexity of testing~$\D$ in the \DHOm\ in terms
of the sample complexity of testing~$\D$ in the standard model.
This leaves open the question of whether
the query complexity of testing~$\D$ in the \DHOm\
can be similarly upper-bounded in terms
of the sample complexity of testing~$\D$ in the \DHOm,
which may be lower than
the sample complexity of testing~$\D$ in the standard model.
A possible avenue towards establishing such a result is
resolving positively the following open problem.

\BO{\em(Preservation of distances under a random relabeling):}
\label{closed-mapping:open}
Suppose that $\D$ is a property of distributions
over $n$-bit strings that is closed under mapping.
Is it the case that if~$X$ is $\e$-far from $\D$,
then, with high probability over the choice of
a random bijection $\pi:\bitset^n\to\bitset^n$,
it holds that $\pi(X)$ is $\Omega(\e)$-far from $\D$?
We stress that the distances here are according to Definition~\ref{dist:def}
and that the hidden constant in the $\Omega$-notation is universal.
\EO
A positive answer to Problem~\ref{closed-mapping:open}
would allow to convert a tester for $\D$ in the \DHOm\
into one that only considers the collision pattern among
the samples. Specifically, given a collision pattern
among $s$ samples, the latter tester will generate at random
a sequence of $s$ samples that fits the given collision pattern,
and invoke the original tester on this sequence of samples.
In such a case, we can apply the strategy used
in the proof of Theorem~\ref{closed-mapping:ithm} to
the resulting tester.

We were able to establish a positive answer
to Problem~\ref{closed-mapping:open} in the special case that
the support of $X$ has size at most $2^{(0.5-\Omega(1))\cdot n}$.
In fact, in that case, we prove a stronger result
(where, for simplicity, 0.49 stands for $0.5-\Omega(1)$).

\BP{\em(A partial answer to Problem~\ref{closed-mapping:open}):}
\label{closed-mapping:prop}
Suppose that $\D$ is a property of distributions
over $n$-bit strings that is closed under mapping,
and that $X$ has support size at most $2^{0.49n}$.
Then, if $X$ is $\e$-far from $\D$ in total variation distance,
then, with high probability over the choice of
a random bijection $\pi:\bitset^n\to\bitset^n$,
it holds that $\pi(X)$ is $\Omega(\e)$-far from $\D$
according to Definition~\ref{dist:def}.
\EP
Some restriction on the support size of $X$ is essential:
Consider $X$ such that $\prob[X\!=\!1^n]=3/4$
and $\prob[X\!=\!0x']=2^{-(n+1)}$ for every $x'\in\bitset^{n-1}$.
Then, $X$ is $\Omega(1)$-far from being $2^n$-grained
in total variation distance,
although (by Observation~\ref{triviality:obs})
it is $o(1)$-close to being $2^n$-grained
(according to Definition~\ref{dist:def}).
\medskip


\BPF
The key observation is that, for some constant $\delta>0$,
with high probability over the choice of
a random bijection $\pi:\bitset^n\to\bitset^n$,
it holds that the elements in the support of $\pi(X)$
are at relative Hamming distance at least $\delta$.
Fixing any such $\pi$, we let $C$ denote the support of $\pi(X)$
and note that $\min_{w\neq w'\in C}\{\Delta_H(w,w')\}\geq\delta$.
Assuming that $X'=\pi(X)$ is $\e'$-close to~$\D$
{\em according to Definition~\ref{dist:def}},
we shall show that $X'$ is $\frac{2}{\delta}\cdot\e'$-close to $\D$
{\em in total variation distance}.
(It follows that $X=\pi^{-1}(X')$ is $\frac{2}{\delta}\cdot\e'$-close to $\D$
{\em in total variation distance}.)

Specifically, we consider a distribution $Y$ in $\D$ such that $X'$
is $\e'$-close to $Y$ according to Definition~\ref{dist:def},
and show that a related distribution $Y'$ that is also in $\D$
is $\frac{2}{\delta}\cdot\e'$-close to $X'$ in total variation distance.
In particular, we shall replace $Y$ by the distribution $Y'$
of the strings in $C$ that are closest to $Y$.

\Bcm{\em(The effect of correction to the closest element of $C$):}
\label{closed-mapping:clm}
Suppose that $X'$ is supported on a set $C$
such that $\min_{w\neq w'\in C}\{\Delta_H(w,w')\}\geq\delta$,
and that $Y$ is $\e'$-close to $X'$
according to Definition~\ref{dist:def}.
Then, $Y'=\DEC(Y)$ is $\frac{2}{\delta}\cdot\e'$-close to $X'$
in total variation distance,
where $\DEC(y)$ denotes a string in $C$ that is closest to $y$.
\Ecm
Recalling that in our application $Y$ is in $\D$,
it follows that $\DEC(Y)$ is in $\D$,
since $\D$ is closed under mapping.
Hence, $X'$ is $\frac{2}{\delta}\cdot\e'$-close to $\D$.
\smallskip

\Bpf
Intuitively, replacing $Y$ by $\DEC(Y)$ may increase the distance from $X'$
according to Definition~\ref{dist:def},
but not too much (i.e., for every $x'\in C$, it holds
that $\Delta_H(x',\DEC(y))\leq2\cdot\Delta_H(x',y)$).%
\footnote{If $x'\neq\DEC(y)$,
then $\Delta_H(y,\DEC(y))\leq\Delta_H(y,x')$.}
%
The key observation is that the distance of $Y'=\DEC(Y)$ to $X'$
(according to Definition~\ref{dist:def})
is due solely to strings that are at Hamming distance at least $\delta$.
This implies that the total variation distance
between $Y'$ and $X'$ is at least a $\delta$ fraction of the
distance between $Y'$ and $X'$ according to Definition~\ref{dist:def}.
Furthermore, we shall show that the total variation distance
between $Y'$ and $X'$ is at least a $\delta/2$ fraction of the
distance between $Y$ and $X'$ according to Definition~\ref{dist:def}.
The actual proof follow.

For $w_{x',y}$'s as in Definition~\ref{dist:def}
(i.e., the minimum sequence of non-negative numbers
that satisfies $\sum_{y} w_{x',y} = \prob[X'\!=\!x']$
and $\sum_{x'} w_{x',y} = \prob[Y\!=\!y]$),
the claim's hypothesis means that
\begin{equation}\label{closed-mapping:hypo:eq}
\sum_{x',y\in\bitset^n} w_{x',y}\cdot \Delta_H(x',y)\leq\e'.
\end{equation}
Recall that $w_{x',y}>0$ only if $x'\in C$,
and that $\DEC(y)$ denote a string in $C$ that is closest to $y$.
Hence,
\begin{eqnarray}
\label{closed-mapping:eq0a}
\lefteqn{\sum_{x',y\in\bitset^n} w_{x',y}\cdot \Delta_H(x',y)} \\
\nonumber
&=& \sum_{y\in\bitset^n} w_{\DEC(y),y}\cdot \Delta_H(\DEC(y),y)
+\sum_{~~y\in\bitset^n~~} \sum_{x'\in C\setminus\{\DEC(y)\}}
       w_{x',y}\cdot \Delta_H(x',y) \\
&\geq& \sum_{~~y\in\bitset^n~~} \sum_{x'\in C\setminus\{\DEC(y)\}}
       w_{x',y}\cdot \Delta_H(x',y).
\label{closed-mapping:eq0b}
\end{eqnarray}
Using $\Delta_H(x',y)\geq\Delta_H(x',\DEC(y))-\Delta_H(\DEC(y),y)$
and $\Delta_H(x',y)\geq\Delta_H(\DEC(y),y)$ for any $x'\in C$,
we have $\Delta_H(x',y)\geq\Delta_H(x',\DEC(y))/2$,
whereas $\Delta_H(x',\DEC(y))\geq\delta$ (for $x'\in C\setminus\{\DEC(y)\}$).
Combining this with \eqrefs{closed-mapping:eq0a}{closed-mapping:eq0b},
we get
\begin{equation}\label{closed-mapping:eq00}
\sum_{x',y\in\bitset^n} w_{x',y}\cdot \Delta_H(x',y)
\geq \sum_{~~y\in\bitset^n~~} \sum_{x'\in C\setminus\{\DEC(y)\}}
        w_{x',y}\cdot \frac{\delta}{2}
\end{equation}
whereas combining \eqref{closed-mapping:hypo:eq}
and \eqref{closed-mapping:eq00}, we get

\begin{equation}\label{closed-mapping:eq1}
\sum_{~~y\in\bitset^n~~} \sum_{x'\in C\setminus\{\DEC(y)\}}
       w_{x',y}\cdot \frac{\delta}{2}\;\leq\;\e'.
\end{equation}
Next, we observe that
the total variation distance between $X'$ and $\DEC(Y)$
is upper-bounded by
\begin{equation}\label{closed-mapping:eq2}
\sum_{~~y'\in C~~} \sum_{x'\in C\setminus\{y'\}}
  \sum_{~~y:\DEC(y)=y'~~} w_{x',y}
  = \sum_{~~y\in\bitset^n~~} \sum_{x'\in C\setminus\{\DEC(y)\}}w_{x',y},
\end{equation}
since the l.h.s of \eqref{closed-mapping:eq2}
represents the cost of a possible way
of transforming $X'$ into $\DEC(Y)$.%
\footnote{That is, we map to $y'$ the mass of $X'$ that was mapped
to $\{y:\DEC(y)=y'\}$
(i.e., $\sum_{x'\in C\setminus\{y'\}}\sum_{y:\DEC(y)=y'} w_{x',y}$),
where we pay one unit per each $x'\neq y'$.
Recall that the total variation distance equals
the earth mover's distance with respect to the inequality function
(i.e., $\InEq(x',y')=1$ if $x'\neq y$ and $\InEq(y',y')=0$).}
%
Combining \eqref{closed-mapping:eq1} and \eqref{closed-mapping:eq2},
it follows that $X'$ is $\frac{2}{\delta}\cdot\e'$-close
to $Y'=\DEC(Y)$ in total variation distance.
\Epf

\myparagraph{Conclusion.}
Recall that we have assumed that $X'$ is $\e'$-closed to $Y$
(according to Definition~\ref{dist:def}), which is in $\D$,
and that $\DEC(Y)$ is in $\D$ (because $\D$ is closed under mapping).
Using Claim~\ref{closed-mapping:clm},
it follows that $X'$ is $\frac{2}{\delta}\cdot\e'$-close
to $\D$ in total variation distance.
Hence, $X=\pi^{-1}(X')$ is $\frac{2}{\delta}\cdot\e'$-close
to $\D$ in total variation distance, since $\D$ is label invariant.%
\footnote{Indeed, unlike Definition~\ref{dist:def},
the total variation distance (between a pair of distributions)
is preserved under relabeling.}
The proposition follows, because
if $X$ is $\e$-far from $\D$ in total variation distance,
then $X'$ must be $\delta\e/2$-far from $\D$
according to Definition~\ref{dist:def}.
\EPF

\subsection{On standard testing of equality of distributions}
\label{std-equality:apdx}
Recall that when proving the furthermore clause
of Theorem~\ref{equal-distributions:thm} we use the fact
that the equality tester (for the standard model) of~\cite{DK}
works also when only the support size of one of
the tested distributions is upper-bounded.%
\footnote{This is not a generic claim regarding any such tester.
Consider, for example, a modification of any tester such that
the modified algorithm always accepts if all samples that
are taken from one of the distributions are distinct.
The latter event is unlikely when the support size of both distributions
is smaller than the square of the number of samples,
but is extremely likely if one distribution is uniform
over a sufficiently large set (i.e., much larger than
the square of the number of samples).\label{std-equality:fn}}
Here, we provide more details about the proof of this fact.

\BT{\em(Testing equality of two unknown distributions, revised):}
Suppose that $X$ and~$Y$ are distributed over $U$
and that one of them has support size at most $n$.
Then, distinguishing between the case that $X\equiv Y$
and the case that $X$ is $\e$-far from $Y$
{\rm(in total variation distance)}
is possible in time $O(\max(n^{2/3}/\e^{4/3},{\sqrt n}/\e^2))$.
\ET
This is a generalization of~\cite[Thm.~11.24]{G:pt},
which originates in~\cite{DK}, where the special case
mandates that $U=[n]$ (and in that case the said
algorithm is a tester of equality).
\medskip

\BPF
We follow the presentation of~\cite[Sec.~11.2--11.3]{G:pt},
and observe that the support size is only used in the proof
of~\cite[Cor.~11.21]{G:pt}, when upper-bounding the total variation
distance between two distributions by the norm-2 of their difference,
whereas essentially the same upper bound holds also if only the support
of one of the distributions is upper-bounded. Details follow.

Our starting point is~\cite[Alg.~11.17]{G:pt},
which is stated as referring to distributions over $[n]$
but can be restated as referring to distributions over $U$.
Recall that the actions of this algorithm only
depend on the $s$ samples it obtains from each distribution,
whereas $s$ is a free parameter.
The same holds with respect to the analysis of
this algorithm as an ${\cal L}_2$-distance approximator,
which is provided in~\cite[Thm.~11.20]{G:pt}.

The key point is that the analysis of~\cite[Alg.~11.17]{G:pt}
as a very crude ${\cal L}_1$-distance approximator,
provided in~\cite[Cor.~11.21]{G:pt},
remains valid under the relaxed hypothesis
(i.e., when only one of the two distributions
is guaranteed to have support size at most $n$).
This is because this upper bound (on the support size)
is only used when upper-bounding the norm-1
(of the difference between the two distributions)
by the norm-2 of the same difference.
We observe that we only lose a factor of two when
performing the argument on the smaller of the two supports,
because at least half of the norm-1 of the difference
is due to this smaller support.
Specifically, let $p:S\to[0,1]$ be the probability function
representing one distribution and $q:U\to[0,1]$ be the function
representing the other distribution, where $S\subseteq U$.
Then,
\begin{eqnarray*}
\sum_{i\in U}|p(i)-q(i)|
&=& 2\cdot\sum_{i\in U:p(i)>q(i)}|p(i)-q(i)| \\
&\leq& 2\cdot\sum_{i\in S}|p(i)-q(i)| \\
&\leq& 2\cdot\sqrt{|S|}\cdot\left(\sum_{i\in S}|p(i)-q(i)|^2\right)^{1/2} \\
&\leq& 2\cdot\sqrt{|S|}\cdot\left(\sum_{i\in U}|p(i)-q(i)|^2\right)^{1/2}
\end{eqnarray*}
where the first inequality is
due to $\{i\!\in\!U\!:\!p(i)>q(i)\}\subseteq\{i\!\in\!U\!:\!p(i)>0\}=S$.
(Indeed, the first and last inequalities are the place
where we go beyond the original proof of~\cite[Cor.~11.21]{G:pt}.)
Hence, $\|p-q\|_1\leq2\sqrt{|S|}\cdot\|p-q\|_2$,
where $|S|\leq n$ by our hypothesis.
(In the original proof of~\cite[Cor.~11.21]{G:pt},
which refers to $p,q:[n]\to[0,1]$,
one gets $\|p-q\|_1\leq\sqrt{n}\cdot\|p-q\|_2$,
but the difference is immaterial.)

Next, we note that~\cite[Cor.~11.22(2)]{G:pt}
remains valid under the relaxed hypothesis
(i.e., when only one of the two distributions
is guaranteed to have support size at most $n$).%
\footnote{We mention that~\cite[Cor.~11.22(1)]{G:pt} also remains valid
(even when both distributions have support of unbounded size),
but it is not used here.}
We stress that this result will only be used
when $\beta\geq n^{-1/2}$ (as presumed in the original text).

Lastly, we turn to~\cite[Alg.~11.24]{G:pt},
which is stated as referring to distributions over $[n]$
but can be restated as referring to distributions over $U$,
while making $n$ a free parameter (just as~$m$ in the original text).
When analyzing this algorithm, we let $n$ denote an upper bound
on the size of the support of one of the two distributions,
and apply the revised~\cite[Cor.~11.22(2)]{G:pt}
(which holds in this case).
Using $m=\min(n^{2/3}/\e^{4/3},n)$
(as in the original text), the current claim follows
(analogously to establishing~\cite[Thm.~11.26]{G:pt}).
\EPF



\end{document}